\documentstyle[mncite,psfig]{mn}

\newcommand{\kms}{km\,s$^{-1}$}
\newcommand{\arcdot}{{$^{\prime\prime}$\hspace*{-0.6ex}.}}
\newcommand{\lesssim}{\raisebox{-0.6ex}{$\,\stackrel
{\raisebox{-.2ex}{$\textstyle <$}}{\sim}\,$}}
\newcommand{\gtrsim}{\raisebox{-0.6ex}{$\,\stackrel
{\raisebox{-.2ex}{$\textstyle >$}}{\sim}\,$}}

\title[VLT spectropolarimetry of powerful radio galaxies]{VLT
spectropolarimetry of two powerful radio galaxies at z\,$\sim$\,1.4\,: UV
continuum, emission-line properties and the nature of high-redshift dust}

\author[C.\ Sol\'{o}rzano-I\~{n}arrea et al.]
{C. Sol\'{o}rzano-I\~{n}arrea$^{1}$\thanks{E-mail:csi@roe.ac.uk},
P. N. Best$^{1}$, H. J. A. R\"{o}ttgering$^{2}$ and A. Cimatti$^{3}$   \\ 
$^{1}$Institute for Astronomy, University of Edinburgh, Royal Observatory, 
Edinburgh, EH9 3HJ, UK\\ 
$^{2}$Sterrewacht Leiden, Postbus 9513, 2300 RA Leiden, The Netherlands \\
$^{3}$Istituto Nazionale di Astrofisica, Osservatorio Astrofisico di
Arcetri, Largo E. Fermi 5, 50125 Firenze, Italy  }

\begin{document}
\maketitle

\begin{abstract}
Deep spectropolarimetric observations, obtained with the Very Large
Telescope (VLT), are presented for two powerful radio galaxies, 0850--206
(z=1.3373) and 1303+091 (z=1.4093). These observations cover the
rest-frame wavelength range $\sim$\,1450\,--\,3750\,\AA. New radio
observations and continuum images of the same sources are also presented.
These galaxies are the first two observed from a complete sample of nine
radio sources with redshifts in the range 1.3\,$\le$\,z\,$\le$\,1.5
(selected from the equatorial sample of powerful radio sources of Best,
R{\"o}ttgering \& Lehnert\nocite{best99b}), as part of a project aimed to
investigate the multi-component nature of the UV continuum in radio
galaxies and, in particular, any variations of the continuum properties
with the radio source age.

The larger radio source of the two, 0850--206, presents a high continuum
fractional polarization, averaging 17 per cent across the observed
wavelength range and reaching 24 per cent at rest-frame wavelengths of
$\lesssim$\,2000\,\AA. The smaller radio source, 1303+091, shows a lower
continuum polarization, averaging 8 per cent and rising to 11 per cent
for rest-frame wavelengths $\gtrsim$\,3000\,\AA. For both galaxies, the
position angle of the electric vector is generally constant with
wavelength and within $\sim$\,15$^{\circ}$ of perpendicular to the radio
axis. Both their total flux spectra and polarized flux spectra reveal the
2200\,\AA \ dust feature, and comparison with dust scattering models
suggests that the composition of the dust in these galaxies is similar to
that of Galactic dust.  In 0850--206, scattered quasar radiation dominates
the UV continuum emission, with the nebular continuum accounting for no
more than $\sim$\,22 per cent and no requirement for any additional
emission component such as emission from young stars. By contrast, in
1303+091, unpolarized radiation could be a major constituent of the UV
continuum emission, with starlight accounting for up to $\sim$\,50 per
cent and the nebular continuum accounting for $\sim$\,11 per cent.

The emission-line properties of the galaxies are also studied from their
total intensity spectra. Comparison of the measured emission-line ratios
with both shock- and photo-ionization models shows that the nuclear and
extended gas in these galaxies is mainly photoionized by the central
active nucleus.
 
\end{abstract}


\section{Introduction}

The galaxies associated with powerful extragalactic radio sources are
uniquely important for understanding the physics of active galactic nuclei
(AGN) and for studying the relationship between radio source activity and
the properties of the host galaxy and its environment.  Powerful distant
radio galaxies are thought to be the progenitors of present day, giant
ellipticals (e.g. \pcite{best98,mclure2000}). However, compared with
normal elliptical galaxies, powerful radio galaxies at z$\gtrsim$0.6 show
enhanced optical/UV continuum and line emission, which is generally
elongated and aligned along the radio axis (e.g. \pcite{mccarthy87}).

The emission-line structures usually extend large distances from the
nucleus (5\,$\sim$\,100 kpc; e.g. \pcite{tadhunter86,baum88}); the
properties of these extended emission-line regions (EELR) can be greatly
influenced by shocks resulting from interactions between the radio source
structures and the interstellar/intergalactic medium (ISM/IGM)
(e.g. \pcite{clark98,carmen01,carmen03}). Interestingly, the emission-line
properties of a sample of z$\sim$1 radio galaxies have been found to
evolve strongly as the radio source passes through the host galaxy
\cite{best2000b}: large radio sources ($\gtrsim$150 kpc) show quiescent
`rotation-dominated' velocity profiles and ionization states in agreement
with AGN-illumination, while smaller radio sources present highly
distorted kinematic profiles and overall ionizations consistent with being
shock-dominated.

As for the optical/UV continuum excess observed in radio galaxies, several
different mechanisms are known to contribute to it
(e.g. \pcite{tadhunter2002}), but the relative extent of their
contributions remains uncertain. A popular early interpretation for this
UV excess was recent star formation, induced by the passage of the radio
jet through~the~ISM of the host galaxy or by merger events linked to the
radio source triggering (e.g. \pcite{rees89}). Jet-induced star formation
has been shown to be feasible in numerical simulations
(e.g. \pcite{mellema2002}). Also, stellar absorption features
characteristic of OB stars have been observed in\,the
radio\,galaxy 4C41.17 at z=3.8 (Dey et al.\,1997)\nocite{dey97}.

The detection of polarized continuum and polarized broad permitted
emission lines in some radio galaxies
(e.g. \pcite{antonu-mill85,di-serego89,cimatti93}) suggested that light
emitted anisotropically by a hidden quasar nucleus and scattered towards
us by dust or electrons in the ISM of the radio galaxy
\cite{tadhunter88,fabian89} also makes an important contribution to the UV
light. This is favoured by orientation-based unified schemes for radio
sources \cite{barthel89}, according to which radio galaxies and quasars
are drawn from the same parent population but viewed from different angles
to the line of sight, with the AGN in radio galaxies obscured by a
surrounding dusty torus.  Spectropolarimetric studies of a small number of
radio galaxies at z$\sim$1 have shown that the polarized emission is
spatially extended, with the electric vector oriented perpendicular to the
UV continuum emission at all wavelengths, and that, while the permitted
MgII\,2800 emission line is observed to be broad in polarized light, none
of the narrow lines shows significant polarization
(e.g. \pcite{cimatti96}, 1997)\nocite{cimatti97}. These results clearly
show that the origin of the polarization is scattered quasar light.  More
recent spectropolarimetric studies of a sample of radio galaxies at
z$\sim$2.5 \cite{vernet2001} have shed light on the nature of the
scattering material. These studies show indications of a continuum upturn
beyond 2200\,\AA \ in the composite spectrum of the galaxies in the
sample, which is interpreted as a possible detection of the 2200\,\AA \
dust feature, indicating that the scattering medium is dust. However,
caution must be taken because, due to the high redshift of their sample,
those wavelengths are redshifted into a region of strong sky lines.

Together with the young stellar and scattered AGN components, other
processes known to contribute to the UV continuum excess in some sources
are nebular continuum emitted by the extended emission-line gas
\cite{dickson95} and direct AGN light \cite{shaw95}.  

As with the emission-line gas, some properties of the continuum emission
in radio galaxies are observed to vary with radio source size: smaller
sources show strings of bright knots aligned along the radio axis, and
larger sources present more compact nebulae with fewer bright components
\cite{best96}.  Do the contributions of the different components to the UV
excess also evolve over the radio source lifetime?  For instance, a large
variation is seen in the strengths, and to a lesser extent colours, of the
polarized emission in radio galaxies. Is this linked to the radio source
size?  Such questions cannot be answered by existing spectropolarimetric
data, because targets were generally selected on the basis of previously
detected polarization in imaging observations, an interesting UV
morphology or an ultra-steep radio spectrum, making the samples studied
biased or incomplete.

To remedy this and address many of the issues raised above, such as the
nature of the scatterers at high redshift and the variation with radio
source size of the different contributions to the UV excess, we have begun
a programme to make deep spectropolarimetric observations of a {\it
complete} sample of nine radio galaxies with redshifts in the range
1.3\,$\le$\,z\,$\le$\,1.5 and a wide range in radio sizes.  In this paper
we present the observations of the first two galaxies from the sample.

The paper is organized as follows. In Section 2, the selection criteria of
our sample are described. Section 3 contains the details of the
observations, data reduction and analysis. The results are presented in
Sections 4 and 5. Discussion of the results follows in Section 6. Summary
and conclusions are provided in Section 7.  Throughout this paper values
of the cosmological parameters of H$_{0}$\,=\,65\,km\,s$^{-1}$\,Mpc$^{-1}$,
$\Omega_{\rm M}$\,=\,0.3 and $\Omega_{\Lambda}$\,=\,0.7 are assumed.

\section{Sample selection}

The sources of our full sample comprise all radio galaxies from the
equatorial sample of powerful radio sources of
\scite{best99b}\footnote{Following the additional identifications and
redshifts of \scite{best2000c} and \scite{best2003}, this radio source
sample is now 100 per cent spectroscopically complete.}, with redshifts in
the range 1.3\,$\le$\,z\,$\le$\,1.5.  The redshift range was chosen to be
low enough to allow good S/N, but also sufficiently high to ensure that
the rest-frame wavelengths studied ($\sim$\,1400\,--\,3800\,\AA) allowed
to search for features in the dust scattering cross-section around
2200\,\AA, as well as stellar absorption lines, such as SiIII\,1417,
SV\,1502 and NIV\,1720. Further, this wavelength range covers the
HeII\,1640 emission line, which can be used as a replacement for the
Balmer lines to derive the amount of nebular continuum emission.  Our
sample also includes sources with a wide range of radio sizes, from $\sim$
10 to 400\,kpc, in order to investigate variations of the emission-line
and continuum properties with the radio source age.

The results obtained from the first two sources observed from our sample
are presented in this paper. Properties of these two sources are listed in
Table~\ref{proptab}.

\section{Observations, data reduction and analysis}

\begin{table}
\begin{center}
\begin{tabular}{ccccc}\hline
{\bf Source}&{\bf z}&{\bf D$_{\rm rad}$}&{\bf PA$_{\rm rad}$}&{\bf E$_{\rm B-V}$}\\
 & & (kpc) & (deg) & (mag)\\ \hline
0850$-$206 & 1.3373$\pm$0.0018 & 118 & 29 & 0.242  \\
1303$+$091 & 1.4093$\pm$0.0012 & 73 & 131 & 0.030  \\ \hline
\end{tabular}
\caption[]{\label{proptab}Properties of the two radio galaxies from our
sample presented in this paper. The redshift of each source, derived from
the data presented in this paper, is given in column 2. Column 3 gives the
projected extent of the radio source along the radio axis
\cite{best99b}. The position angle of the radio axis for each source (Best
et al. 1999) is listed in column 4. The last column gives the colour
excess E$_{\rm B-V}$ for each source taken from NASA/IPAC Extragalactic
Database (NED).}
\end{center}
\end{table}

\subsection{VLT observations}

Optical imaging and spectropolarimetric observations were carried out on
the 8.2-m ESO Very Large Telescope (VLT) Antu (UT1) at Paranal (Chile) on
the nights of 2000 January 31 and February 1. The imaging spectrograph
FORS1 was used in both direct imaging (IMG) and multi-object
spectropolarimetry (PMOS) modes together with the Tek 2048$\times$2048
CCD, which provides a spatial scale of 0.2 arcsec\,pixel$^{-1}$.
Observations were made under good seeing conditions, averaging 0.5 arcsec
and 0.7 arcsec for the two nights, respectively. The first night was
clear, but part of the second night was non-photometric. Details of the
observations are presented in Tables~\ref{vltimgtab} and
\ref{vltspecpoltab}.

\subsubsection{Optical imaging}

Imaging observations were made through the broad-band filters V\_BESS+35
for 0850--206, and I\_BESS+37 for both 0850--206 and 1303+091.

The reduction of the imaging data was performed using standard tasks
within the {\footnotesize IRAF} software package, following the usual
steps.  The bias was first subtracted and then the images were
flat-fielded using sky flats. Photometric calibration was achieved by
using observations of the photometric standard stars PG\,0231+051 and
PG\,1323--086 taken during the nights. The Johnson photometric magnitudes,
measured through 2- and 4-arcsec diameter apertures, are given in
Table~\ref{vltimgtab}.

\subsubsection{Optical spectropolarimetry}

The grism GRIS\_300V was used for the spectropolarimetric observations,
providing a spectral dispersion of approximately 2.6 \AA\,pixel$^{-1}$.
For the source 0850--206 a 1.4-arcsec slit was oriented along the axis
defined by the position of optical objects, which is misaligned by
approximately 35$^{\circ}$ with the radio axis [see
Fig.~\ref{vradalpha0850} (left)]. In the case of 1303+091, a 1-arcsec slit
was oriented parallel to the radio axis [see Fig.~\ref{iradalpha1303}
(left)]. Note that the imaging spectrograph FORS has a compensator that
corrects for atmospheric dispersion. Four sets of observations were made
for each source (two sets on each night), each set consisting of four
exposures with the half-wave plate position angle rotated to 0$^{\circ}$,
22$^{\circ}$\hspace*{-.8ex}.5, 45$^{\circ}$ and
67$^{\circ}$\hspace*{-.8ex}.5, consecutively.  0850--206 was observed in
three sets of four 1350\,s exposures and one set of four 900\,s exposures,
with a total integration time of 5.5\,hr. 1303+091 was observed in two
sets of four 1350\,s, one set of four 900\,s and one set of three 1150\,s
(the 67$^{\circ}$\hspace*{-.8ex}.5 exposure was lost due to instrument
problems), making a total integration time of~nearly 5\,hr. The last two
sets of 1303+091 were carried out in non-photometric conditions.

\begin{table}
\begin{center}
\begin{tabular}{ccccccc}\hline
{\bf Source} &{\bf Filter}&{\bf $\lambda_{\rm cen}$} &{\bf t$_{\rm exp}$} &
\multicolumn{2}{c}{\bf magnitude}& {\bf Airm.} \\
  &  & (\AA) & (s) & 2$^{\prime\prime}$ & 4$^{\prime\prime}$  & \\ \hline
0850--206 & V\_BESS & 5540 & 300 & 22.98 & \hspace*{-0.15cm}22.73 & 1.024 \\
0850--206 & I\_BESS & 7680 & 300 & 21.79 & \hspace*{-0.15cm}21.39 & 1.018 \\
1303+091  & I\_BESS & 7680 & 300 & 21.83 & \hspace*{-0.15cm}21.40 & 1.724 \\ \hline
\end{tabular}
\caption{\label{vltimgtab}Log of the VLT optical imaging observations.
Photometric magnitudes, measured through 2- and 4-arcsec diameter
apertures, are also given.}
\end{center}
\end{table}

The reduction of the spectropolarimetric data was performed using
{\footnotesize IRAF} in the following way.  The frames were first
bias-subtracted and then flat-fielded using lamp flats taken through the
same optics used in the source and star frames. After removing cosmic-ray
events, the different exposures for the same half-wave plate positions
were combined for each source. In the case of 1303+091, appropriate
weights and scales were applied to account for the non-photometric
conditions in which this source was observed during the second night.
One-dimensional spectra, with identical aperture widths for the ordinary
(o) and extraordinary (e) rays for each of the four half-wave plate
orientations, were then extracted.  The wavelength calibration was
performed using He, HgCd and Ar arc lamps. After that, o- and e-rays were
resampled to the same linear spectral dispersion, in order to calculate
the polarization over identical spectral bins. The data were then binned,
and all eight one-dimensional spectra were combined to determine the
Stokes parameters I, Q and U for each galaxy, using the procedures
described in Vernet (2001;\nocite{thesisvernet2001} see also
\pcite{vernet2001}), which are based on the method outlined in
\scite{cohen95}. Observations of polarized (HD245310 and HD111579) and
unpolarized (HD64299 and HD94851) standard stars
\cite{turnshek90,schmidt92} were taken in order to check and calibrate the
instrumental polarization and the position angle zero-point offset.  Flux
calibration was provided by observations of the spectrophotometric
standard stars Feige\,24 and G163--50. The atmospheric extinction
correction was achieved by using the extinction curve of La Silla
Observatory, since no suitable data for Paranal are available.

\begin{table*}
\begin{center}
\begin{tabular}{ccccccccc}\hline
 
{\bf Source}&{\bf t$_{\rm exp}$}&{\bf Slit PA}&{\bf Grism}&{\bf $\lambda$
 Range}&{\bf Pixel Scale}&{\bf Resolution}&{\bf Slit Width}&{\bf Airmass}\\
 & (s) & (deg.) &  & (\AA) & (arcsec/\AA) & (\AA) & (arcsec) & \\ \hline

0850--206 & 19800 & 175 & GRIS\_300V & 3460 -- 8792 & 0.2/2.607 &
14.66$\pm$0.10 & 1.4 & 1.19 \\
1303+091  & 17850 & 131 & GRIS\_300V & 3281 -- 8583 & 0.2/2.593 &
11.27$\pm$0.07 & 1.0 & 1.36 \\ \hline
\end{tabular}
\caption{\label{vltspecpoltab}Log of the VLT optical spectropolarimetric
observations. The spectral resolution given is that of the observed frame.
}
\end{center}
\end{table*}

No second-order filter was used in these observations, but instead the
polarized standard star HD245310 was observed both with and without the
GG435 filter, which blocks the light below 4200\,\AA, to estimate the
effects of any second-order blue light.  Residual second-order light,
which only affects the spectral region with
$\lambda$\,$\gtrsim$\,7000\,\AA, was found to be very low.  In addition,
the continuum polarization of the star HD245310 was derived for both sets
of observations (with and without filter), and it was found that it did
not vary in any significant way. Thus, any contamination by second-order
blue light does not affect the conclusions of this paper.

For the analysis of the total flux spectra, the two-dimensional frames of
the o- and e-rays were first corrected for any distortion along the
spectral direction, using the Starlink {\footnotesize FIGARO} package, and
then summed. After that, one-dimensional spectra were extracted, and then
analysed using the Starlink {\footnotesize DIPSO} spectral analysis
package. Emission-line fluxes, velocity shifts and linewidths (FWHM) were
obtained by Gaussian fitting of the emission-line profiles. The measured
linewidths were corrected for the spectral resolution of the instrument
via Gaussian deconvolution.  The instrumental resolution was measured from
night-sky lines and arc lines, and is listed in Table~\ref{vltspecpoltab}.

The total intensity spectra, emission-line fluxes and ratios have been
corrected for Galactic reddening using the extinction curve from
\scite{seaton79} and values of colour excess ${\rm E_{B-V}}$ from
NASA/IPAC Extragalactic Database (NED; see Table~\ref{proptab}).

\subsection{Radio observations}

\begin{table}
\begin{center}
\begin{tabular}{ccccc}\hline

{\bf Source}&{\bf Freq.}&{\bf Array}&{\bf Obs. Date}&{\bf t$_{\rm exp}$}\\
 & (MHz) & (Config.) & (d/m/y) & (min) \\ \hline

0850--206 & 8460 & A & 09/12/2000 & 56 \\
          &      & B & 20/05/2001 & 16 \\
0850--206 & 4710 & A & 09/12/2000 & 38 \\
          &      & B & 20/05/2001 & 15 \\
1303+091 & 8460 & A & 11/12/2001 & 55 \\
         &      & B & 20/05/2001 & 16 \\
1303+091 & 4710 & A & 11/12/2001 & 33 \\
         &      & B & 20/05/2001 & 15 \\ \hline
\end{tabular}
\caption{\label{raddatatab}Log of the VLA radio observations.}
\end{center}
\end{table}

New radio observations of all nine sources in the spectropolarimetric
sample were carried out using the Very Large Array radio
interferometer. Each source was observed using the A-array configuration
at both 8.5 and 4.7\,GHz frequencies, mapping both the total intensity and
polarization properties. Sources larger than 10 arcsec in angular extent
were additionally mapped using the B-array configuration in order to
provide sensitivity to the extended structures. Full details of all of
these observations will be provided elsewhere (Best et al., in
prep.). Here, only the total intensity data for 0850--206 and 1303+091 are
considered, to compare the radio properties with the spectropolarimetric
and imaging data and to use the radio spectral index information to locate
the radio cores.

Details of the observations of these two sources are provided in
Table~\ref{raddatatab}.  Data reduction for both sources followed the
standard procedures of {\footnotesize CLEAN}ing and self-calibration
within the {\footnotesize AIPS} package (e.g. see \pcite{best99a}).
8.5\,GHz total intensity maps were made at full angular resolution
($\approx$~0.3 arcsec), with rms noise levels of 22\,$\mu$Jy. Maps were
also made at lower angular resolutions matching those of the 4.7\,GHz data
(0.45 arcsec for 1303+091; 0.67 by 0.45 arcsec for 0850--206), by
weighting down the outer regions of the UV-plane. These matched-resolution
8.5\,GHz and 4.7\,GHz maps were aligned and combined to produce a radio
spectral index map of each source.

\subsection{Registration of the radio and optical images}

\begin{figure*}
\centerline{ \psfig{figure=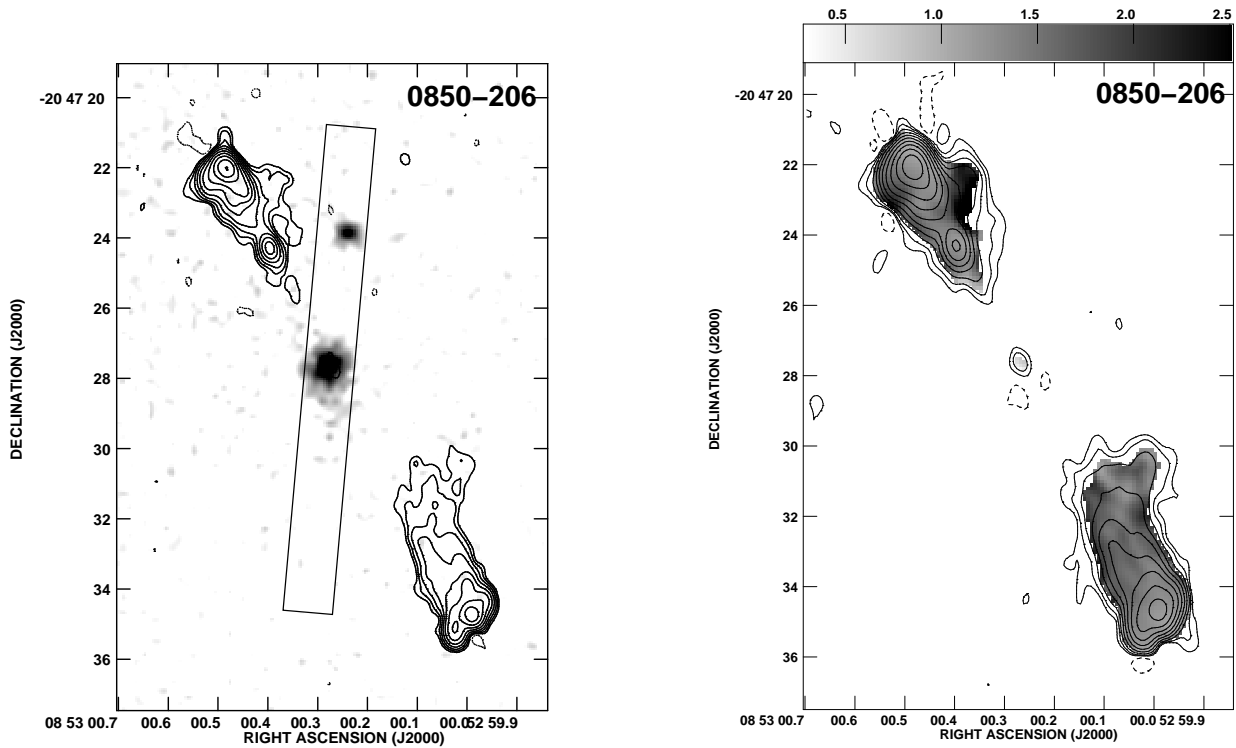,width=16.5cm}}
\caption[]{Left: V-band image of 0850--206 overlaid with contours of the
radio emission at 8.5\,GHz. The values of the radio contours are
1$\times$10$^{-4}$ Jy\,beam$^{-1}$\,$\times$\,(--1, 1, 2, 4, 8, 16, 32,
64, 128, 256, 512). The position of the slit is also indicated on the left
plot (the length of the slit is arbitrary).  Right: Radio spectral index,
$\alpha$ (greyscale, defined as $S_{\nu}$\,$\propto$\,$\nu^{-\alpha}$),
between 8.5 and 4.7\,GHz for 0850--206.  Overlaid are contours of radio
emission at 4.7\,GHz, with contour levels of
1.7$\times$10$^{-4}$\,Jy\,beam$^{-1}$\,$\times$\,(--1, 1, 2, 4, 8, 16, 32,
64, 128, 256, 512). The bright radio knot at RA\,=\,08h\,53m\,00.27s
Dec\,=\,--20$^{\circ}$\,47$^{\prime}$\,27.6$^{\prime\prime}$ has the
flattest spectral index, $\alpha$\,$\approx$\,0.4 which is typical of
radio cores at these redshifts (e.g. \pcite{athreya97}).}
\label{vradalpha0850}
\end{figure*}

The optical and radio maps of the two galaxies were aligned by assuming
that the continuum centroid of the optical emission was coincident with
the location of the radio core, adopted to be the radio knot with the
flattest spectral index [see Figs.~\ref{vradalpha0850} (right) and
\ref{iradalpha1303} (right)]. As a check, astrometry was also performed on
the optical images by using several unsaturated stars present in the
observed fields and in the APM \cite{maddox90} and US Naval Observatory
catalogues. 

\section{0850--206}

\subsection{The continuum emission}
\label{res:0850con}

Fig.~\ref{vradalpha0850} (left) presents the VLT image of the radio galaxy
0850--206 in the V band, overlaid with contours of the radio emission at
8.5 GHz.  The position of the slit for the spectropolarimetric
observations is also shown in the figure. The central wavelength of the
V-filter in the rest frame of the galaxy was 2370 \AA, with a bandwidth of
440 \AA \ (rest-frame). 

\begin{figure*}
\centerline{ \psfig{figure=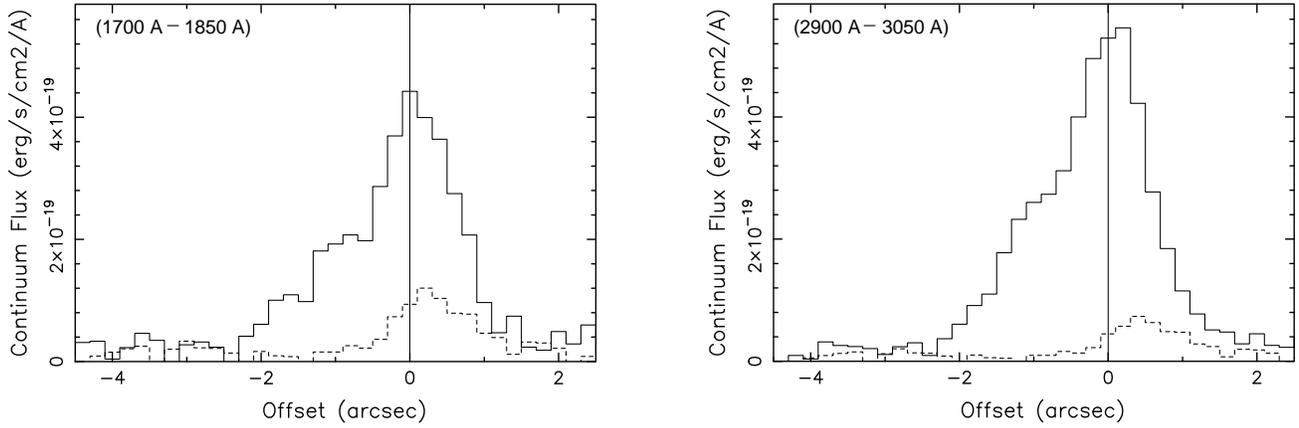,width=17.5cm}}
\caption[]{Spatial profiles of the line-free continuum emission (solid
line) along the slit for 0850--206, between 1700\,--\,1850 \AA \
rest-frame (left) and between 2900\,--\,3050 \AA \ rest-frame (right). The
dashed line shows the nebular continuum profile for the same wavelength
ranges (see Section~\ref{diss:uv_origin}). South is to the left.}
\label{cont0850}
\end{figure*}

Fig.~\ref{cont0850} shows the spatial profile of the line-free continuum
emission (solid line) along the slit for 0850--206, for the rest-frame
wavelength ranges 1700\,--\,1850 \AA \ (left) and 2900\,--\,3050 \AA \
(right).  It appears that the total continuum emission in the UV end is
slightly weaker than that at longer wavelengths.

\subsubsection{Polarization of the continuum emission}

\begin{figure*}
\centerline{ \psfig{figure=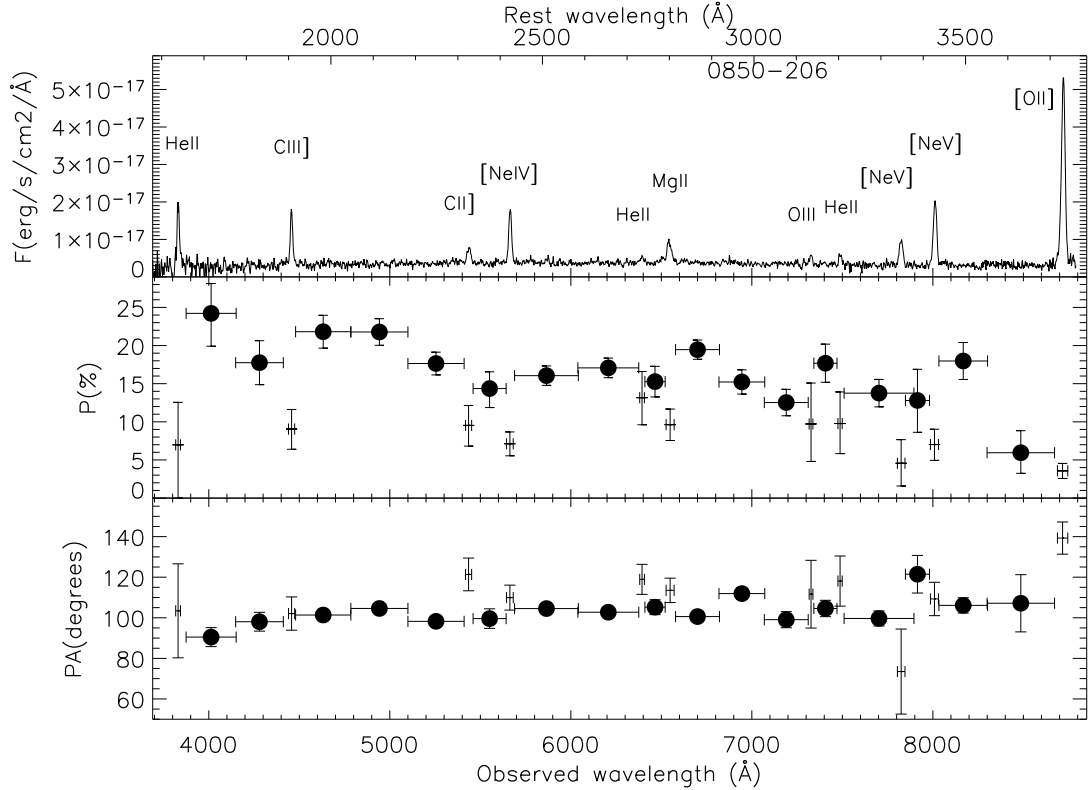,width=15cm}}
\caption[]{Spectral and polarization properties of the radio galaxy
0850--206.  From top to bottom: the total flux spectrum, the observed
fractional polarization and the position angle of the electric vector.
Filled circles and crosses indicate respectively continuum bins and
emission line bins with their underlying continuum.}
\label{pol0850}
\end{figure*}

The spectropolarimetric results of 0850--206 are presented in
Fig.~\ref{pol0850}. From top to bottom the following are plotted: the total
flux spectrum, the percentage polarization and the position angle of the
electric vector. A 2 arcsec-wide ($\sim$\,18\,kpc) aperture centred on the
continuum centroid was used for the extraction of 1D spectra for the o-
and e-rays.

The continuum fractional polarization of this source is observed to
increase gradually towards shorter wavelengths, with an average value of
17 per cent across the observed wavelength range of the spectrum. The
highest polarization is measured in the bin longward of HeII\,1640, with a
value of 24.2$^{+3.9}_{-4.3}$per cent. Note that the polarization measured
in the bin shortward of [OII]3727, with a value of 6 per cent, may be
affected by noise due to the strong sky lines in that spectral region.

The position angle of the electric vector is generally constant with
wavelength, with an average value of 103$^{\circ}$, which is within
$\sim$\,15$^{\circ}$ perpendicular to the radio axis (PA$_{\rm
rad}$\,$\sim$\,29$^{\circ}$, defined by the position of the two most
distant radio hotspots).  For shorter rest-frame wavelengths than
1900\,\AA, the position angle seems to decrease slightly down to a value
of $\sim$\,90$^{\circ}$. Previous studies of high-z radio galaxies show a
better perpendicularity between the electric vector and the UV continuum
axis than with the radio axis (e.g. \pcite{di-serego93};
\pcite{cimatti93}, 1994\nocite{cimatti94};
\pcite{di-serego96,vernet2001}). There are no rest-frame UV images of
0850--206 available to measure the UV axis with precision, but the UV
image published in this paper [\,Fig.~\ref{vradalpha0850} (left)],
although not very deep, suggests that the UV axis is
12$^{\circ}$$\pm$5$^{\circ}$, which is perpendicular to the electric
vector orientation. It should also be noted that previous studies of low-z
narrow-line radio galaxies have shown that when there is a hidden nucleus
and a polarization fan, the measured position angle of the electric vector
can depend on the slit orientation \cite{cohen99}.

\subsection{The line emission}

\subsubsection{Emission-line structure}
\label{res:line0850}

\begin{figure*}
\centerline{ \psfig{figure=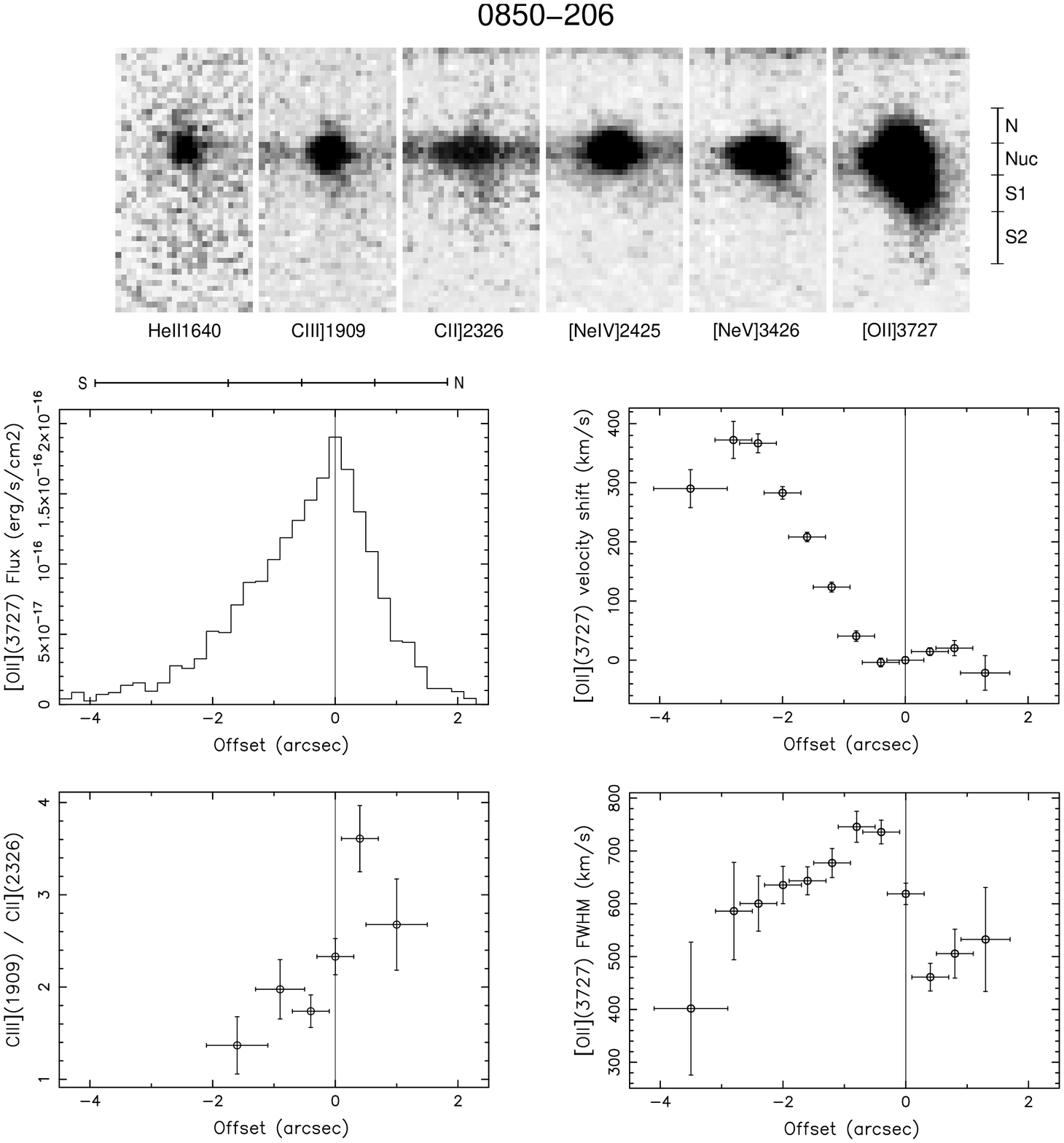,width=17.5cm}}
\caption[]{{\bf Top panel:} 2D spectra of the main emission lines for
0850--206. Each spectrum covers 9.6 arcsec along the slit, and 30 \AA \
(rest-frame) in the spectral direction. The regions defined along the slit
are indicated by the solid line at the right of the panel. {\bf Bottom
panel:} Emission-line spatial analysis along the slit for 0850--206. Top
left: Variation of the [OII]3727 flux. The spatial regions defined along
the slit are indicated by a solid line at the top of the plot. Bottom
left: Spatial variation in the CIII]1909\,/\,CII]2326 line ratio. Top
right: Spatial variation in the [OII]3727 velocity centroids. Bottom
right: Spatial variation of the [OII]3727 linewidth. South is to the left
in the four plots.}
\label{linestudy0850}
\end{figure*}

2D spectra of some of the main emission lines observed for 0850--206 are
shown in the top panel of Fig.~\ref{linestudy0850}. These are HeII\,1640,
CIII]1909, CII]2326, [NeIV]2425, [NeV]3426 and [OII]3727. The spatial
variation of the flux of the brightest emission line ([OII] in this case)
along the slit was determined and it is presented in the same figure (bottom
panel -- top left).  The spatial profile was derived by first extracting a
spatial slice perpendicular to the dispersion direction and centred on the
[OII] emission line, with a spectral width of 60\,\AA. Then, to correct
for the continuum contamination, a spatial slice extracted from a
line-free region of the spectrum, adjacent to the emission line and scaled
to have the same spectral width as that of the emission-line slice, was
subtracted from the first one.

The detected [OII] emission extends approximately 6 arcsec
($\sim$\,54\,kpc) along the slit. The line emission distribution is
observed to be asymmetric with respect to the nucleus, extending further
to the south than to the north.

\begin{figure}
\centerline{ \psfig{figure=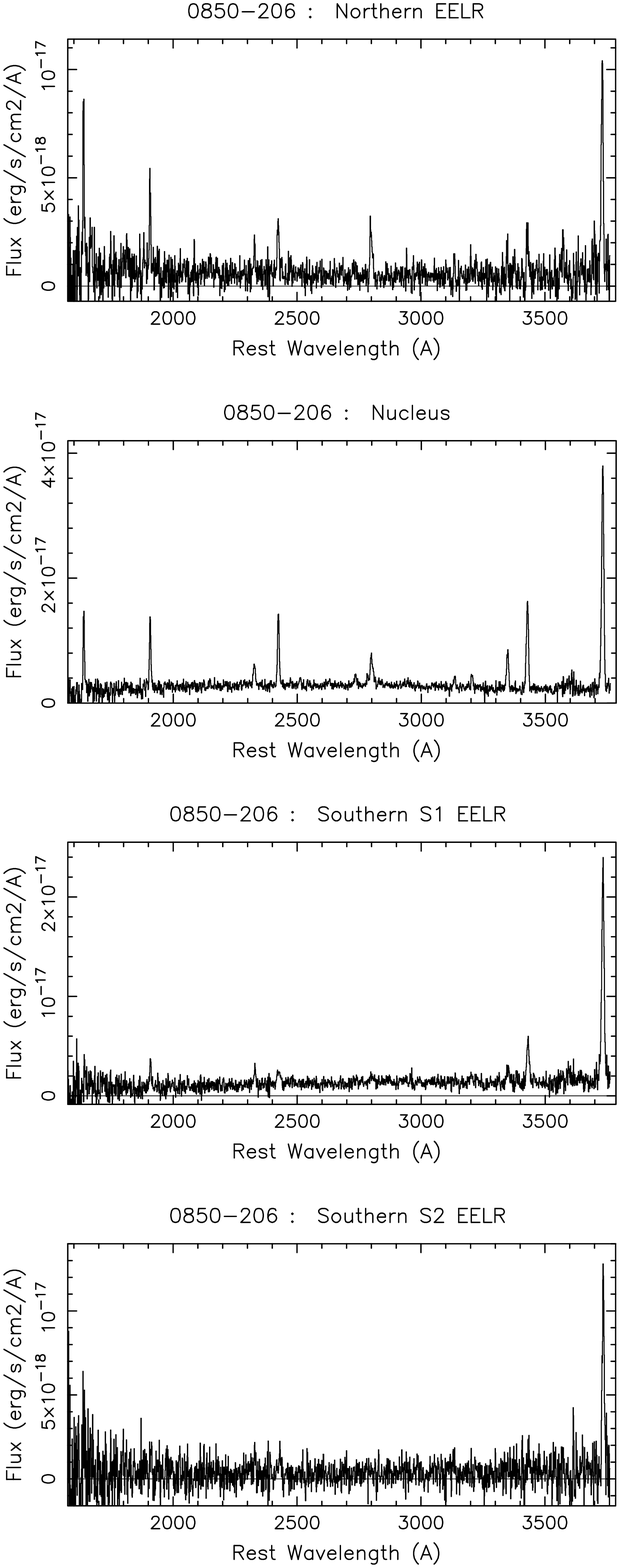,width=8.35cm}}
\caption[]{Integrated spectra for the different emission-line regions in
0850--206 defined along the slit: northern EELR (top; 1.2$\times$1.4
arcsec$^{2}$ aperture centred at 1.1 arcsec N of the continuum centroid),
nucleus (middle top; 1.2$\times$1.4 arcsec$^{2}$ aperture centred at 0.1
arcsec S of the continuum centroid), southern EELR `S1' (middle bottom;
1.2$\times$1.4 arcsec$^{2}$ aperture centred at 1.3 arcsec S of the
continuum centroid) and southern EELR `S2' (bottom; 2.2$\times$1.4
arcsec$^{2}$ aperture centred at 3.0 arcsec S of the continuum centroid). }
\label{0850specreg}
\end{figure}

\subsubsection{Emission-line kinematics}
\label{res:kin0850}

Kinematic information of the emission-line gas in 0850--206 was obtained
by fitting single Gaussians to the profile of [OII]3727, which was the
emission line with the highest signal-to-noise ratio in the observed
spectrum. To try to match the seeing of the observations, one-dimensional
spectra with an aperture of 3 pixels (0.6 arcsec) and centred in steps of
2 pixels were extracted from the 2D total intensity spectrum of the
galaxy. In the outermost regions of the galaxy the aperture width of the
extracted 1D spectra was greater than 3 pixels in order to increase the
S/N of these regions.
  
The spatial variation along the slit of the velocity centroids and
linewidth of the [OII]3727 line for 0850-206 are shown in
Fig.~\ref{linestudy0850} (bottom panel -- top right, bottom right). 

The velocity shifts, which are referred to the velocity of the
emission-line gas at the continuum centroid, are observed to vary smoothly
along the position of the slit, with an overall velocity amplitude of
$\Delta$$v$\,$\sim$\,450 \kms. The velocity curve is, however, highly
asymmetric with respect to the nucleus; while the emission-line gas at the
north end, about 1 arcsec from the centre, appears to have a velocity
comparable to that of the nucleus, the gas at $\sim$\,3 arcsec to the
south of the nucleus is redshifted by about +400 \kms.

The instrumentally-corrected linewidths (FWHM) of the [OII] line are found
to range between $\sim$\,300 and 800 \kms \ along the location of the
slit. The broadest emission lines, with FWHM\,$\sim$\,750\,\kms, are
located about 1 arcsec south of the nucleus. The nucleus itself presents
[OII] linewiths of FWHM\,$\sim$\,620 \kms.

\subsubsection{Emission-line spectra}
\label{res:spec0850}

The total intensity spectrum of 0850--206 (corresponding to an aperture of
2 arcsec centred on the nucleus) is shown in Fig.~\ref{pol0850}
(top). Also, integrated spectra of four different regions defined along
the slit (nuclear region, southernmost `S2', southern `S1' and northern
EELR) are presented in Fig.~\ref{0850specreg} (see the caption of the
figure for details of the apertures).  The emission-line fluxes normalized
to the CII]\,2326 flux for each region are listed in
Table~\ref{tabflux0850cor}. CII]\,2326 was chosen for the normalization of
the line fluxes because it is a common line in the observed spectra of
both 0850--206 and 1303+091, whose emission is also detectable in the
extended regions of the galaxies.

The spatial variation of the CIII]1909\,/\,CII]2326 line ratio along the
location of the slit for 0850--206 is shown in Fig.~\ref{linestudy0850}
(bottom panel -- bottom left). The emission-line gas with the highest
ionization state is found to be at $\sim$\,0.5 arcsec north of the
nucleus, with a line ratio CIII]\,/\,CII]\,$\sim$\,3.6. The nucleus itself
presents a moderate ionization state in comparison with the rest of the
regions observed. The gas showing the lowest ionization, with
CIII]\,/\,CII]\,$\sim$\,1.3, is located at $\sim$\,2 arcsec south of the
nucleus, which coincides with a region of relatively broad emission lines
(FWHM$_{\rm [OII]}$\,$\sim$\,660 \kms).

\subsubsection{Polarization of the emission lines}
\label{res:polline0850}

\begin{table*}
\begin{tabular}{lcccc}\hline
\vspace*{0.7ex}
\hspace*{0.5cm}{\bf 0850--206}&{\bf S2\,EELR}&{\bf $\!$S1\,EELR}&{\bf Nucleus}&{\bf N EELR}\\ 
{\bf Line} & (3\arcdot0 S)  &(1\arcdot3 S) & (0\arcdot1 S) & (0\arcdot9 N) \\ \hline

HeII\,1640 \dotfill& $<$3.28 & $<$1.06 & 2.56$\pm$0.21& 5.22$\pm$0.95 \\
NIII$]$\,1749 \dotfill& --- & --- & 0.32$\pm$0.13 & 1.07$\pm$0.46 \\
SiIII$]$\,1882 \dotfill& --- & --- & 0.18$\pm$0.06 & --- \\
SiIII$]$\,1892 \dotfill& --- & --- & 0.25$\pm$0.06 & --- \\
CIII$]$\,1909 \dotfill& 1.17$\pm$0.43 & 1.37$\pm$0.26 & 2.46$\pm$0.17 &
2.92$\pm$0.52 \\
CII$]$\,2326 \dotfill& 1.00$\pm$0.23 & 1.00$\pm$0.13 & 1.00$\pm$0.06 & 1.00$\pm$0.16 \\
$[$NeIV$]$\,2425 \dotfill& 1.21$\pm$0.46& 1.50$\pm$0.28 & 2.95$\pm$0.18 & 2.19$\pm$0.39 \\
$[$OII$]$\,2472 \dotfill& --- & --- & 0.23$\pm$0.05 & --- \\
$[$MgVII$]$\,2510 \dotfill& --- & --- & 0.33$\pm$0.06 & --- \\
$[$MgVII$]$\,2629 \dotfill& --- & --- & 0.21$\pm$0.06 & --- \\
HeII\,2733 \dotfill& --- & --- & 0.53$\pm$0.07 & --- \\
$[$MgV$]$\,2783 \dotfill& --- & --- & 0.28$\pm$0.05 & --- \\
MgII\,2800 \dotfill& --- & --- & 1.92$\pm$0.14 & 2.53$\pm$0.44 \\
HeI\,2829 \dotfill& --- & --- & 0.14$\pm$0.04 & --- \\
OIII\,3133 \dotfill& --- & --- & 0.42$\pm$0.06 & 0.64$\pm$0.17 \\
HeII\,3203 \dotfill& --- & 0.55$\pm$0.16 & 0.52$\pm$0.06 & --- \\
$[$NeV$]$\,3346 \dotfill& --- & 1.20$\pm$0.24 & 1.80$\pm$0.13 & 1.12$\pm$0.28 \\
$[$NeV$]$\,3426 \dotfill& $<$1.08 & 3.69$\pm$0.53 & 4.28$\pm$0.26 & 1.96$\pm$0.41 \\
$[$OII$]$\,3727 \dotfill& 10.88$\pm$2.56 & 18.43$\pm$2.44 &
11.50$\pm$0.69 & 9.54$\pm$1.55 \\ \hline

CII$]$\,2326 flux   & & & & \\
(10$^{-17}$ erg\,s$^{-1}$\,cm$^{-2}$) & 2.42$\pm$0.55 & 3.11$\pm$0.41 & 7.77$\pm$0.45 & 2.25$\pm$0.35 \\ \hline
\end{tabular}
\caption{\label{tabflux0850cor}Emission-line integrated fluxes (corrected
for Galactic reddening and normalized to the CII]2326 flux) for the
different regions in 0850--206.  The details of the apertures are in the
caption of Fig.~\ref{0850specreg}.
The quoted errors correspond to the line fitting errors.}
\end{table*}

In view of the unified schemes for radio sources, it is expected that the
forbidden lines are unpolarized, since they are emitted isotropically in
the narrow-line region outside the obscuring torus (unless resonantly
scattered). On the other hand, the broad permitted lines, emitted in the
broad-line region inside the torus, are expected to be polarized.

Fig.~\ref{pol0850} (middle plot) shows the observed fractional
polarization for continuum and emission-line bins in 0850--206, for an
spatial aperture of 2 arcsec centred on the continuum centroid. It can be
seen that the polarization of the emission lines (with the underlying
continuum) is lower than the adjacent pure continuum bins. After
subtracting the underlying continuum, the polarization of the emission
lines with high enough S/N was measured, obtaining the following: P$_{\rm
CIII]}$\,=\,5.1$^{+3.9}_{-4.6}$ per cent, P$_{\rm [NeIV]}$\,=\,3.7$\pm$2.5
per cent, P$_{\rm [NeV]}$\,=\,4.1$\pm$2.9 per cent and P$_{\rm
[OII]}$\,=\,3.6$\pm$1.1 per cent. The position angles of the electric
vector for these lines range from 112$^{\circ}$ to 135$^{\circ}$.  These
low polarizations (if real) could be due to transmission through
magnetically aligned dust in either the radio galaxy itself or the ISM of
our Galaxy: using the Galactic relation P$_{\rm ISM}$$<$\,9.0\,E$_{\rm
B-V}$ \cite{serkowski75}, the interstellar polarization due to our Galaxy
could be up to $\sim$\,2.2 per cent.  However, the fact that the electric
vector position angles of these lines are on average almost perpendicular
to the radio axis of 0850--206 (PA$_{\rm rad}$\,$\sim$\,29$^{\circ}$),
suggests that the low polarization observed in these emission lines could
be due to scattering or residuals of the continuum subtraction.

The continuum-subtracted polarization of the permitted MgII line was also
measured, obtaining P$_{\rm MgII}$\,=\,4.6$^{+4.0}_{-4.6}$ per cent and an
electric vector position angle of 148$^{\circ}$$\pm$24$^{\circ}$. The
errors in the polarization measurements of this line are bigger and thus
no significant conclusions can be drawn about its polarization properties.
  
\section{1303+091}

\subsection{The continuum emission}
\label{res:1303con}

\begin{figure*}
\centerline{ \psfig{figure=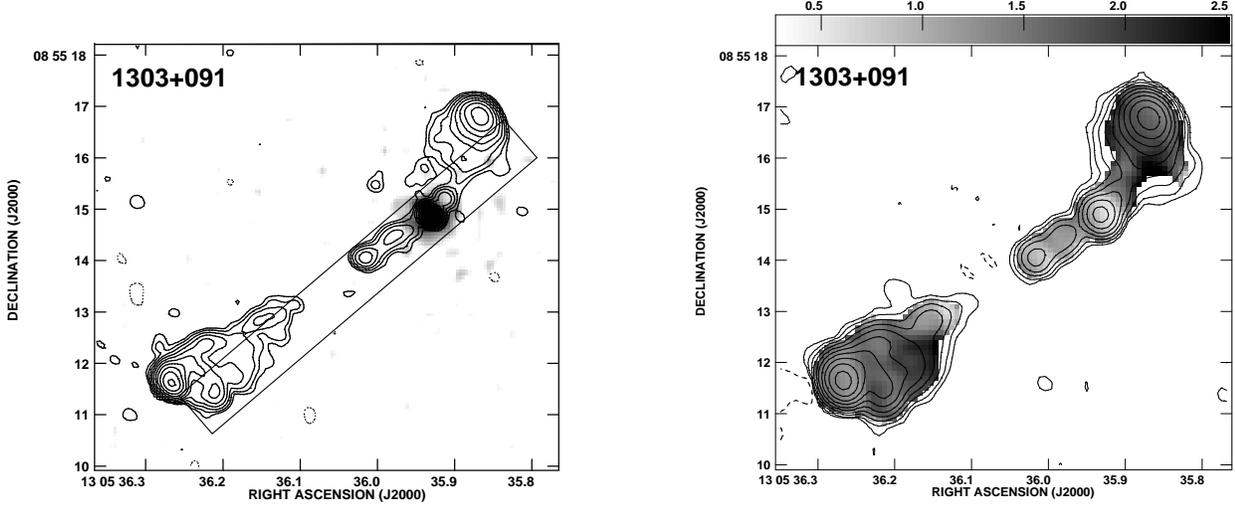,width=16.5cm}}
\caption[]{Left: I-band image of 1303+091 overlaid with contours of the
radio emission at 8.5\,GHz. The values of the radio contours are
7\,$\times$\,10$^{-5}$ Jy\,beam$^{-1}$\,$\times$\,(--1, 1, 2, 4, 8, 16,
32, 64, 128, 256, 512). The position of the slit is also indicated (the
length of the slit is arbitrary). Right: Radio spectral index, $\alpha$
(greyscale, defined as $S_{\nu}$\,$\propto$\,$\nu^{-\alpha}$), between 8.5
and 4.7\,GHz for 1303+091.  Overlaid are contours of radio emission at
4.7\,GHz, with contour levels of
1.5\,$\times$\,10$^{-4}$\,Jy\,beam$^{-1}$\,$\times$\,(--1, 1, 2, 4, 8, 16,
32, 64, 128, 256, 512). The bright radio knot at RA\,=\,13h\,05m\,35.93s
Dec\,=\,+08$^{\circ}$\,55$^{\prime}$\,14.9$^{\prime\prime}$ has the
flattest spectral index, $\alpha$\,$\approx$\,0.5 which is typical of
radio cores at these redshifts (e.g. \pcite{athreya97}).}
\label{iradalpha1303}
\end{figure*}

Fig.~\ref{iradalpha1303} (left) shows the VLT I-band image of the radio
galaxy 1303+091, overlaid with contours of the radio emission at 8.5
GHz. The position of the slit for the spectropolarimetric observations is
indicated in the figure. The central wavelength of the I-filter in the
rest frame of the galaxy was 3188 \AA, with a bandwidth of 573 \AA \
(rest-frame).

Fig.~\ref{cont1303} shows the spatial profile of the line-free continuum
emission (solid line) along the slit for 1303+091, for the rest-frame
wavelength ranges 1700 -- 1850 \AA \ (left) and 2900 -- 3050 \AA \
(right).  It can be seen that, unlike 0850--206, 1303+091 presents a
stronger continuum emission at the UV end than at longer wavelengths.

\subsubsection{Polarization of the continuum emission}

Fig.~\ref{pol1303} presents the spectropolarimetric results of 1303+091,
showing from top to bottom: the total flux spectrum, the percentage
polarization and the position angle of the electric vector. The aperture
used for the extraction of the 1D spectra for the o- and e-rays was 2
arcsec ($\sim$\,18\,kpc) wide and centred on the continuum centroid.

Compared with 0850--206, 1303+091 presents an overall lower continuum
fractional polarization, with an average value of 8 per cent across the
observed wavelength range of the spectrum. The highest polarization is
found in the bin longward of [NeV]3426, with a value of
11.5$^{+5.6}_{-5.2}$ per cent. The polarization then, by contrast with
0850--206, is observed to decrease slightly towards shorter wavelengths. A
dip is found in the bin of $\sim$\,2000\,\AA\ (rest frame) with a value of
4.8$\pm$1.0 per cent.  For shorter wavelengths than $\sim$\,1900\AA, the
polarization is observed to increase again up to a value of
7.5$^{+2.4}_{-2.5}$ per cent.

The position angle of the electric vector is fairly constant with
wavelength, with an average value of 25$^{\circ}$, which is within
$\sim$\,15$^{\circ}$ perpendicular to the radio axis, defined by the
position of the two most distant radio hotspots (PA$_{\rm
rad}$\,$\sim$\,131$^{\circ}$), and within 7$^{\circ}$ perpendicular to the
axis defined by the radio core and the south-eastern inner radio knot
(PA$_{\rm knot}$\,$\sim$\,122$^{\circ}$).  For longer rest frame
wavelengths than $\sim$\,3300\,\AA, the position angle of the electric
vector has values of 64$^{\circ}$$\pm$10$^{\circ}$ and
--5$\pm$14$^{\circ}$ for the two separate bins considered; however these
values are not likely to be significant since that region coincides with
the presence of strong sky emission lines, and thus it is very noisy. No
comparison could be made between the electric vector orientation and the
UV continuum axis because there are no rest-frame UV images available of
1303+091 to derive the UV axis.

\subsection{The line emission}

\subsubsection{Emission-line structure}

2D spectra of some of the main emission lines observed for 1303+091 are
shown in the top panel of Fig.~\ref{linestudy1303}. These are CIV\,1548,
HeII\,1640, CIII]1909, CII]2326, [NeIV]2425 and [NeV]3426. The spatial
profile along the radio axis of the emission line with the highest
signal-to-noise ratio in both nuclear and extended regions (CIII] in this
case) was obtained, and it is presented in the same figure (bottom panel
-- top left).  This profile was obtained in the same way as described for
0850--206 in Section~\ref{res:line0850}, but for 1303+091 the spectral
width of the CIII] slice was 40\,\AA.

The detected CIII] emission extends about 3.5 arcsec ($\sim$\,32 kpc)
along the radio axis. The line emission is asymmetrically distributed with
respect to the nucleus; in addition to the central nuclear peak, there is
a fainter peak at about 1 arcsec ($\sim$\,9 kpc) to the north-west of the
nucleus, close to the radio knot in the north lobe where the radio source
appears to bend.

\subsubsection{Emission-line kinematics}

Kinematic information for the emission-line gas in 1303+091 was obtained
following the same procedure as for 0850--206 (see
Section~\ref{res:kin0850}). In this case, the emission line used for the
analysis was the CIII]\,1909 line, which has the highest S/N in the
observed spectrum and a similar ionization state to [OII]\,3727 used for
0850--206.

The spatial variation of the velocity centroids and linewidth of CIII]
along the radio axis of 1303+091 are presented in Fig.~\ref{linestudy1303}
(bottom panel -- top right, bottom right).

The velocity shifts are referred to the velocity of the emission-line gas
at the continuum centroid. By contrast with 0850--206, the overall
velocity amplitude around the nucleus along the radio axis of 1303+091
changes by $\Delta$$v$\,$\lesssim$\,100 \kms \ over $\sim$\,3 arcsec.

\begin{figure*}
\centerline{ \psfig{figure=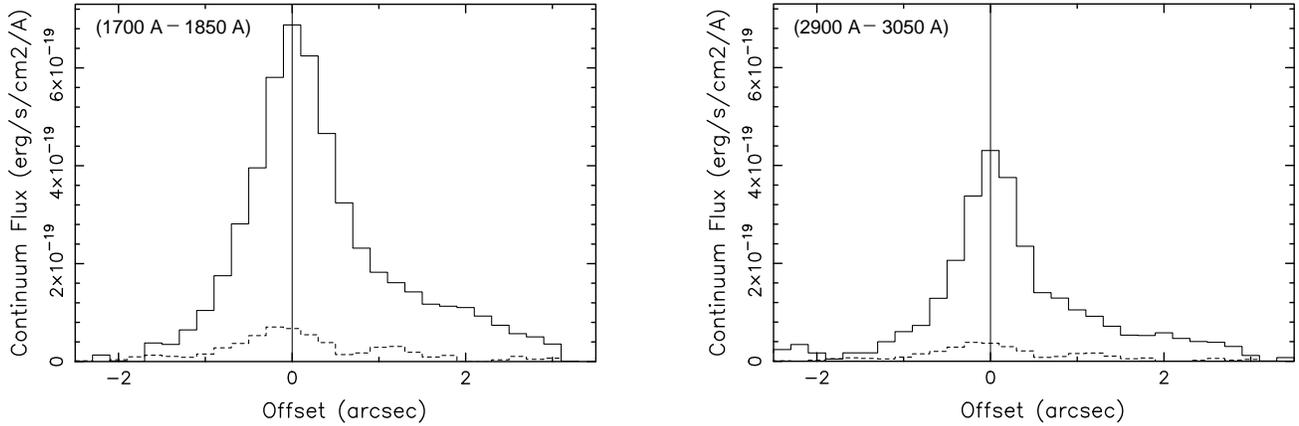,width=17.5cm}}
\caption[]{Spatial profiles of the line-free continuum emission (solid
line) along the radio axis of 1303+091, between 1700\,--\,1850 \AA \
rest-frame (left) and between 2900\,--\,3050 \AA \ rest-frame (right). The
dashed line shows the nebular continuum profile for the same wavelength
ranges (see Section~\ref{diss:uv_origin}). South-east is to the left.}
\label{cont1303}
\end{figure*}

\begin{figure*}
\centerline{ \psfig{figure=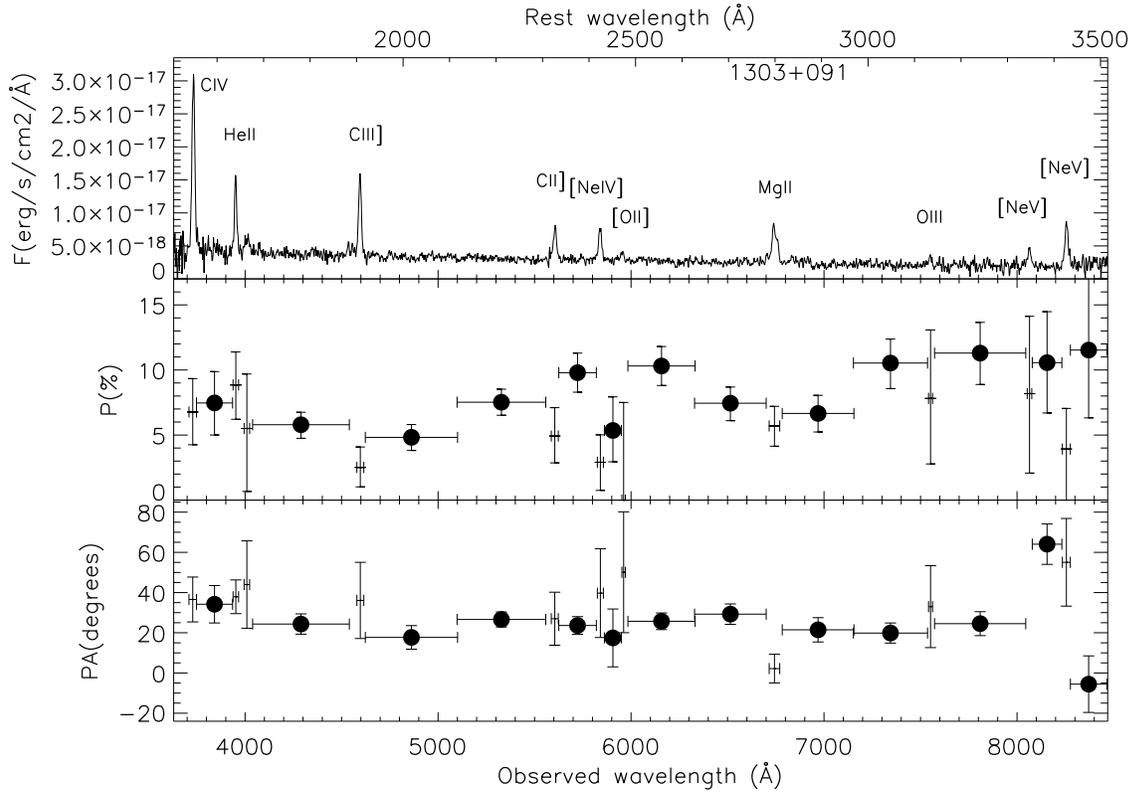,width=15cm}}
\caption[]{Spectral and polarization properties of the radio galaxy
1303+091.  From top to bottom: the total flux spectrum, the observed
fractional polarization and the position angle of the electric
vector. Filled circles and crosses indicate respectively continuum bins
and emission line bins with their underlying continuum.}
\label{pol1303}
\end{figure*}

The CIII]~instrumentally-corrected~linewidths (FWHM) along the radio axis
of 1303+091 are observed to vary between $\sim$\,600 \kms \ at 2 arcsec
north-west of the nucleus and 900\,\kms \ at 1 arcsec south-east of the
nucleus. The CIII] linewidth at the location of the continuum centroid is
found to be FWHM\,$\sim$\,800 \kms. It is interesting that the highest
linewidths are measured at the location of the south-eastern inner radio
knot, and that the narrowest lines are found outside the radio structure
of the northern radio lobe (see Fig.~\ref{iradalpha1303}). This is
expected if the broadening of the emission lines is due to interactions
with the radio structures.

\subsubsection{Emission-line spectra}

\begin{figure}
\centerline{ \psfig{figure=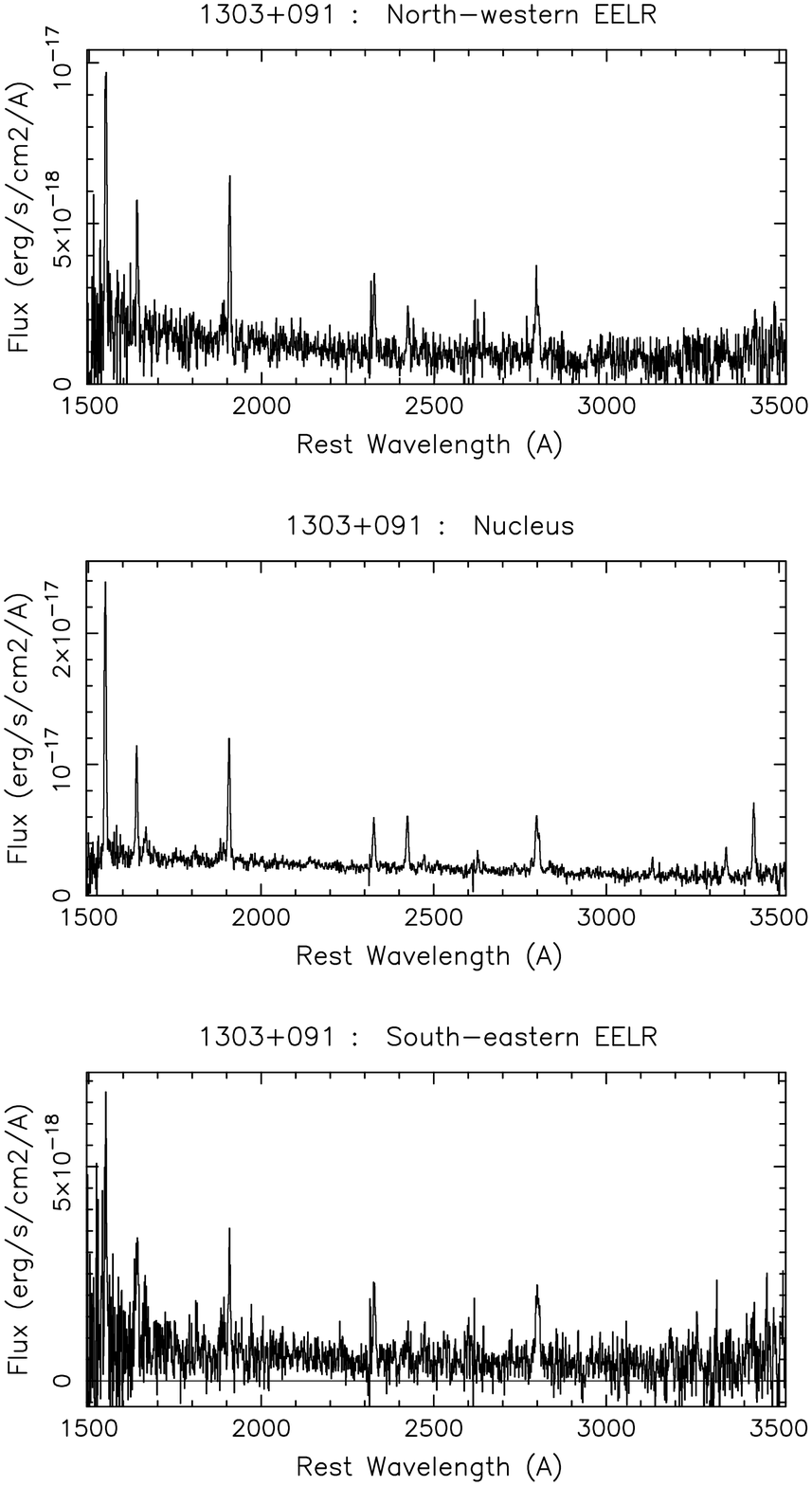,width=8.35cm}}
\caption[]{Integrated spectra for the different emission-line regions in
1303+091 defined along the radio axis: north-western EELR (top;
1.8$\times$1.0 arcsec$^{2}$ aperture centred at 1.4 arcsec NW of the
continuum centroid), nucleus (middle; 1.0$\times$1.0 arcsec$^{2}$ aperture
centred on the continuum centroid) and south-eastern EELR (bottom;
1.6$\times$1.0 arcsec$^{2}$ aperture centred at 1.3 arcsec SE of the
continuum centroid). }
\label{1303specreg}
\end{figure}

The total intensity spectrum of 1303+091 (corresponding to an aperture of
2 arcsec centred on the nucleus) is shown in Fig.~\ref{pol1303} (top). In
addition, integrated spectra of three regions defined along the radio axis
(nuclear region, south-eastern and north-western EELR) are presented in
Fig.~\ref{1303specreg} (see the caption of the figure for details of the
apertures).  The emission-line fluxes normalized to the CII]\,2326 flux
(see Section~\ref{res:spec0850}) for each region are listed in
Table~\ref{tabflux1303cor}.

\begin{figure*}
\centerline{ \psfig{figure=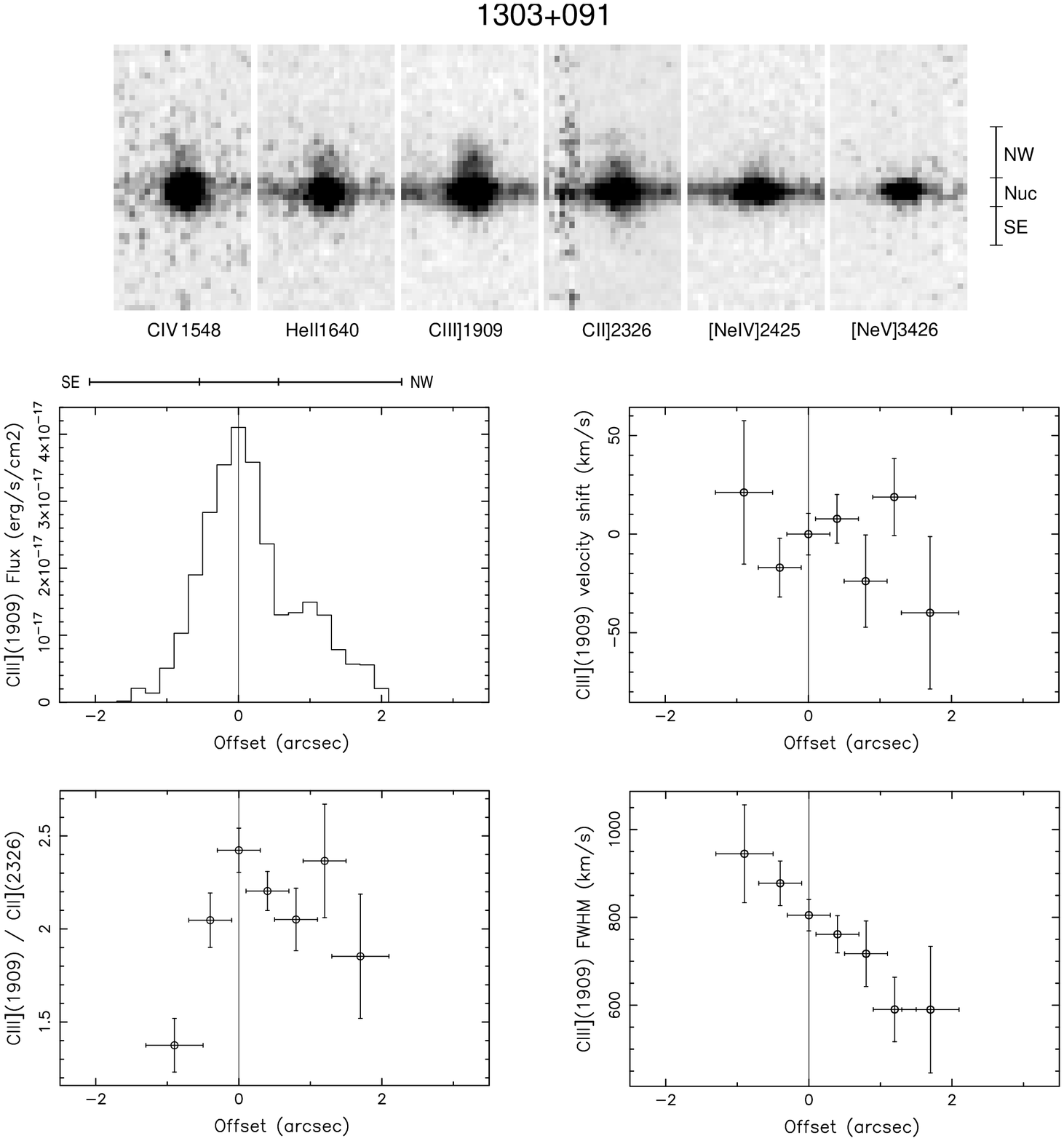,width=17.5cm}}
\caption[]{{\bf Top panel:} 2D spectra of the main emission lines for
1303+091. Each spectrum covers 9.6 arcsec along the radio axis, and 29~\AA
\ (rest-frame) in the spectral direction. The regions defined along the
slit are indicated by the solid line at the right of the panel. {\bf
Bottom panel:} Emission-line spatial analysis along the radio axis of
1303+091. Top left: Spatial variation of the CIII]\,1909 flux. The spatial
regions defined along the slit are indicated by a solid line at the top of
the plot. Bottom left: Spatial variation in the CIII]1909\,/\,CII]2326
line ratio. Top right: Spatial variation in the CIII]\,1909 velocity
centroids. Bottom right: Spatial variation of the CIII]\,1909
linewidth. South-east is to the left in the four plots.}
\label{linestudy1303}
\end{figure*}

The spatial variation of the CIII]1909\,/\,CII]2326 line ratio along the
radio axis of 1303+091 is shown in Fig.~\ref{linestudy1303} (bottom panel
-- bottom left).  The nucleus presents the highest ionization state along
the radio axis, with a line ratio CIII]\,/\,CII]\,$\sim$\,2.5. There may
be a second peak in the ionization state at $\sim$\,1.2 arcsec north-west
of the continuum centroid. The emission-line gas with the lowest
ionization is located at $\sim$\,1 arcsec south-east of the nucleus, with
a ratio CIII]\,/\,CII]\,$\sim$\,1.4. Interestingly, this region of low
ionization presents the broadest emission lines (FWHM$_{\rm
CIII]}$\,$\sim$\,950 \kms), and coincides with the location of a radio
knot (see Fig.~\ref{iradalpha1303}).

\subsubsection{Polarization of the emission lines}

\begin{table}
\hspace*{-0.55cm}
\begin{tabular}{lccc}\hline
\vspace*{0.5ex}
\hspace*{0.5cm}{\bf 1303+091}&{\bf SE\,EELR}&{\bf Nucleus}&{\bf NW\,EELR}\\ 
{\bf Line} & (1\arcdot3 SE) &  & (1\arcdot4 NW) \\ \hline

CIV\,1548 \dotfill  & 2.10$\pm$0.33 & 4.84$\pm$0.20 & 3.06$\pm$0.32 \\
HeII\,1640 \dotfill & 1.59$\pm$0.36 & 1.71$\pm$0.11 & 1.40$\pm$0.24  \\
OIII$]$\,1666 \dotfill & --- & 0.53$\pm$0.12 & --- \\
$[$MgVI$]$\,1806 \dotfill & --- & 0.16$\pm$0.04 & --- \\
SiIII$]$\,1882 \dotfill & --- & 0.22$\pm$0.04 & --- \\
SiIII$]$\,1892 \dotfill & --- & 0.23$\pm$0.04 & --- \\
CIII$]$\,1909 \dotfill & 1.08$\pm$0.13 & 2.28$\pm$0.09 & 1.89$\pm$0.17 \\
NII$]$\,2139 \dotfill & --- & 0.09$\pm$0.02 & --- \\
CII$]$\,2326 \dotfill & 1.00$\pm$0.08 & 1.00$\pm$0.04 & 1.00$\pm$0.08 \\
$[$NeIV$]$\,2425 \dotfill & --- & 1.09$\pm$0.05 & 0.55$\pm$0.09 \\
$[$OII$]$\,2472 \dotfill & --- & 0.20$\pm$0.03 & --- \\
$[$MgVII$]$\,2510 \dotfill & --- & 0.11$\pm$0.02 & --- \\
$[$MgVII$]$\,2629 \dotfill & --- & 0.27$\pm$0.03 & --- \\
HeII\,2733 \dotfill & --- & 0.12$\pm$0.02 & --- \\
$[$MgV$]$\,2783 \dotfill & 0.18$\pm$0.08 & 0.21$\pm$0.04 & 0.30$\pm$0.14 \\
MgII\,2800 \dotfill & 1.39$\pm$0.17 & 1.70$\pm$0.07 & 1.30$\pm$0.17 \\
OIII\,2836 \dotfill & --- & 0.13$\pm$0.02 & 0.23$\pm$0.17 \\
OIII\,3133 \dotfill & --- & 0.26$\pm$0.04 & --- \\
HeII\,3203 \dotfill & --- & 0.14$\pm$0.04 & --- \\
$[$NeV$]$\,3346 \dotfill & --- & 0.54$\pm$0.05 & --- \\
$[$NeV$]$\,3426 \dotfill & --- & 1.45$\pm$0.08 & $<$0.67 \\ \hline

CII$]$\,2326 flux & & & \\
 (10$^{-17}$\,erg\,s$^{-1}$\,cm$^{-2}$)& 3.71$\pm$0.30 & 7.10$\pm$0.25 & 4.32$\pm$0.34 \\ \hline
\end{tabular}
\caption{\label{tabflux1303cor}Emission-line integrated fluxes (corrected
for Galactic reddening and normalized to the CII]2326 flux) for the
different regions in 1303+091. The details of the apertures are in the
caption of Fig.~\ref{1303specreg}.
The quoted errors correspond to the line fitting errors.}
\end{table}

Fig.~\ref{pol1303} (middle plot) shows the observed fractional
polarization for continuum and emission-line bins in 1303+091, for a
spatial aperture of 2 arcsec centred on the continuum centroid. It can be
noticed that the polarization of the emission lines (with the underlying
continuum) is similar or lower than the adjacent pure continuum
bins. After subtracting the underlying continuum, the polarization of the
stronger emission lines was measured. The permitted lines where found to
be significantly polarized, P$_{\rm CIV}$\,=\,9.4$^{+4.3}_{-4.1}$ per
cent, P$_{\rm HeII}$\,=\,11.7$^{+4.4}_{-4.3}$ per cent and P$_{\rm
MgII}$\,=\,5.6$^{+2.6}_{-2.4}$ per cent (note that these values are for
the entire emission line: there was insufficient S/N to attempt to
separate any broad and narrow components). The low Galactic reddening
towards 1303+091 means that the ISM in our Galaxy makes a negligible
contribution to the polarization (P$_{\rm ISM}$$<$\,0.27 per cent;
Serkowski et al. 1975). The position angles of the electric vector for CIV
and HeII are 42$^{\circ}$$\pm$13$^{\circ}$ and
36$^{\circ}$$\pm$11$^{\circ}$, respectively, which are almost
perpendicular to the radio axis of 1303+091 (PA$_{\rm
rad}$\,$\sim$\,131$^{\circ}$), suggesting that the polarization of CIV and
HeII (if real) is dominated by scattering. The electric vector orientation
for MgII is, however, found to be 167$^{\circ}$$\pm$13$^{\circ}$. This
value is far from perpendicular to the radio axis, and it is probably
affected by the low S/N of the MgII line.

The polarization of the stronger forbidden and semi-forbidden lines was
also measured, finding: P$_{\rm CIII]}$\,=\,3.6$^{+2.5}_{-2.4}$ per cent,
P$_{\rm [NeIV]}$\,=\,4.5$^{+4.9}_{-4.5}$ per cent and P$_{\rm
[NeV]}$\,=\,6.7$^{+5.5}_{-6.7}$ per cent, all of which are consistent with
zero.
    
\section{Discussion}

\subsection{The origin of the UV continuum}
\label{diss:uv_origin}

There is evidence that at least four mechanisms contribute to the UV
excess observed in powerful radio galaxies [\,e.g. \scite{tadhunter2002} and
references therein]; these are: scattered AGN light, young stellar
population, nebular continuum and direct AGN light.  In this section the
relative contribution of the different components to the UV continuum of
0850--206 and 1303+091 is investigated.

Direct AGN light is found to contribute significantly to the UV continuum
of a source only when broad permitted lines are detected in the intensity
spectrum. Such sources are likely to be partially obscured, or low
luminosity, quasars and they usually present low levels of polarization of
order $\lesssim$\,4 per cent (e.g. \pcite{stockman79,antonucci84}).  In
the cases of 0850--206 and 1303+091, no broad components are detected in
the profiles of the permitted emission lines in their spectra; also, for
both galaxies the observed polarization is on average $\gtrsim$\,8 per
cent. In addition to this, inspection of the broad-band images of
0850--206 and 1303+091 shows no evidence for a point source in the nucleus
of either galaxy. Direct AGN light is therefore not an important mechanism
in the UV continuum of 0850--206 or 1303+091.  Thus, nebular emission,
young stars and scattered AGN light will be the main mechanisms
contributing to the UV continuum emission of both sources.

\begin{table*}
\begin{center}
\begin{tabular}{ccccccc}\hline

{\bf Source} & {\bf $\lambda$ range} (\AA)& $P_{o}$ (\%) & $P_{obs}$ (\%)&
 $F_{neb}$ (\%)& $F_{star}$ (\%)& $F_{scatt}$ (\%) \\ \hline

0850--206 & 1700\,--\,1850 & 6\,--\,16 & 23$\pm$3 & $\sim$ 22 & $\sim$ 0 & $\sim$ 78 \\
        & 2900\,--\,3050 & 6\,--\,17 & 15.6$\pm$1.3 & $\sim$ 11 & $\sim$ 0 & $\sim$ 89 \\ \hline
1303+091 & 1700\,--\,1850 & 6\,--\,16 & 6.2$\pm$1.2 & $\sim$ 11 & $\sim$ 0\,--\,50 & $\sim$ 39\,--\,89 \\
        & 2900\,--\,3050 & 6\,--\,17 & 10.1$\pm$1.7 & $\sim$ 11 & $\sim$ 0\,--\,29 & $\sim$ 60\,--\,89 \\ \hline

\end{tabular}
\caption[]{\label{poltabboth}Approximate relative contributions of the
nebular continuum $F_{neb}$, stellar population $F_{star}$ and scattered
AGN light $F_{scatt}$ to the total continuum emission of 0850--206 and
1303+091 in the rest wavelength ranges 1700\,--\,1850 \AA \ and
2900\,--\,3050 \AA, evaluated for a 2-arcsec wide aperture centred at the
continuum centroid of the galaxies. Also are given the ranges of intrinsic
continuum polarization $P_{o}$ for dust scattering, assuming a half-cone
opening angle of 45$^{\circ}$ and angles between the cone axis and the
line of sight in the range 50$^{\circ}$--\,90$^{\circ}$ \cite{manzini96},
together with the observed polarization $P_{obs}$ in the two wavelength
ranges.}
\end{center}
\end{table*}

The theoretical nebular continuum for Case B recombination
\cite{osterbrock89} can be derived from the strength of the recombination
lines, H$\beta$ in particular. As the observed wavelength range for
0850--206 and 1303+091 did not cover the H$\beta$ line, the nebular
continuum emission of these galaxies was instead derived by using the
recombination line HeII\,1640 and the average line ratio HeII\,/\,H$\beta$
$\sim$ 3.18 given by \scite{mccarthy93} for radio galaxies.  The spatial
profiles of the nebular continuum emission for 0850--206 and 1303+091 are
shown in Figs.~\ref{cont0850} and \ref{cont1303}, respectively.  To
calculate these profiles, a slice perpendicular to the dispersion
direction, of 40\,\AA \ width in the observed frame, containing the HeII
emission line was extracted from the total two-dimensional spectrum of
each galaxy. Then, adjacent line-free continuum slices were extracted,
combined and scaled to have the same spectral width as the HeII slice. The
combined continuum slice was then subtracted from the HeII slice to
produce a continuum-free HeII spatial profile, which was used to generate
the nebular continuum emission profile (using the task {\footnotesize
NEBCONT} in {\footnotesize DIPSO}). For both sources, the nebular
contribution is generally more important in the extended regions than in
the nucleus. In 0850--206, essentially all of the light south of 2 arcsec
from the nucleus is nebular emission.  For the apertures used in the
polarization analysis, the contribution of the nebular continuum to the
total continuum emission in 0850--206 is found to be $\sim$\,22 per cent
for the rest-frame wavelength range 1700\,--\,1850 \AA, and $\sim$\,11 per
cent for the rest-frame range 2900\,--\,3050 \AA. In the case of 1303+091,
the nebular contribution is $\sim$\,11 per cent for both rest-frame
wavelength ranges 1700\,--\,1850 \AA \ and 2900\,--\,3050 \AA. Thus, at
least $\sim$\,80 per cent of the UV continuum in these galaxies will be a
combination of stellar light and scattered AGN light \footnote{The stellar
light will include both the old stellar population of the galaxy and any
young stellar population. Based on K-magnitudes at these redshifts (K-z
relation; e.g. \pcite{mccarthy93}), the old stellar population is likely
to contribute $\lesssim$\,10 per cent of the light at 3000\,\AA, and will
be negligible at wavelengths $<$\,2500\,\AA.}. It is likely that if the
continuum polarization of the source is low, the contribution of a young
stellar population will be important. On the other hand, if it is the case
that the observed continuum polarization is high, then the scattered
quasar light will probably dominate the UV continuum emission of the
source.

The relative contributions of the continuum components can be roughly
quantified by assuming that the observed fractional polarization $P_{obs}$
is diluted by unpolarized radiation, which is mainly composed of nebular
continuum emission $F_{neb}$ and stellar light $F_{star}$.  Thus we can
use the following equation:

\begin{equation}
P_{obs} = P_{o} \times \frac{F_{scatt}}{F_{scatt} + F_{neb} + F_{star}}
\label{eq:pol}
\end{equation}

\noindent
where $P_{o}$ is the intrinsic undiluted polarization and $F_{scatt}$ is
the scattered AGN flux. Assuming that dust scattering is the dominant
mechanism for 0850--206 and 1303+091 (see Section~\ref{diss:scatterers}
below for a discussion about the scatterers), and taking into account that
\scite{zubko2000} showed that polarization is almost independent of
optical depth, the calculations derived by \scite{manzini96} for optically
thin dust scattering can be used to obtain a good estimate of the
intrinsic polarization of the scattered radiation. Their model uses the
geometry indicated by the unified schemes for radio sources
\cite{barthel89} and assumes that the nuclear radiation is emitted along
two diametrically opposed cones with a half-opening angle of 45$^{\circ}$.
Depending on the angle between the cone axis and the line of sight, the
observed intrinsic polarization for a given wavelength will vary. Using
equation~(\ref{eq:pol}) and assuming that the observer's viewing angle
relative to the cone axis varies between 50$^{\circ}$ and 90$^{\circ}$,
the approximate contributions of the different components to the continuum
emission of 0850--206 and 1303+091 can be estimated for the
rest-wavelengths ranges 1700\,--\,1850\,\AA \ and 2900\,--\,3050\,\AA. The
results are listed in Table~\ref{poltabboth}.

\begin{figure*}
\centerline{ \psfig{figure=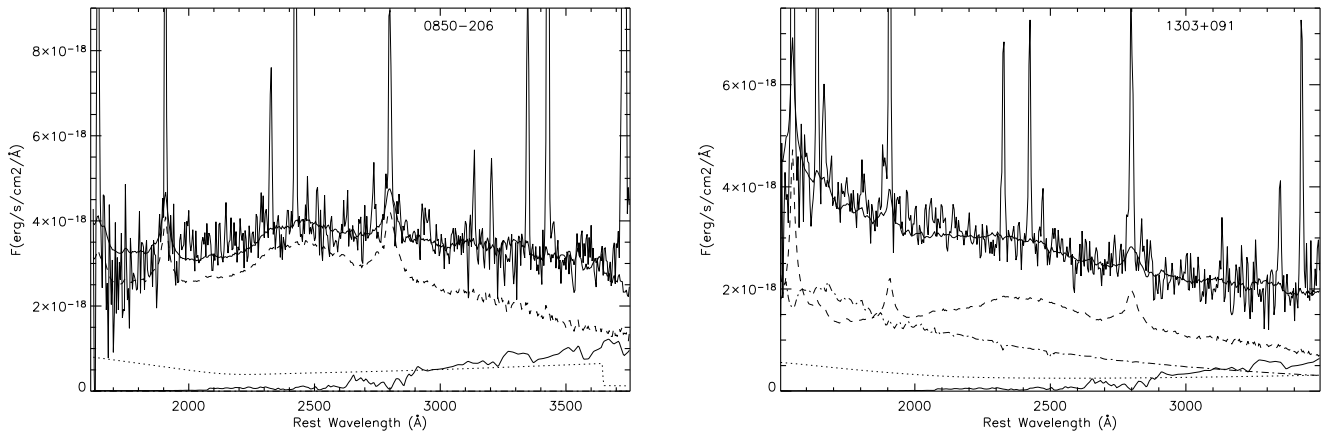,width=17.7cm}}
\caption[]{Four-component fit to the spectra of 0850--206 (left) and
1303+091 (right). The components of the fit include: nebular continuum
(dotted line), dust-scattered radio-loud quasar light (dashed line), young
stellar population (dot-dash-dot line) and old stellar population (thin
solid line). The total fit is indicated by a thick solid line. Both
the 0850--206 and 1303+091 spectra have been smoothed into 4\,\AA
\ (rest-frame) bins.}
\label{sedfit}
\end{figure*}

In the case of 0850--206 the contribution of young stellar population is
negligible or null, and the scattered AGN light dominates the continuum
emission, with a relative contribution of $\sim$\,80\,--\,90 per cent
depending on the nebular contribution for the different wavelength
ranges. Based on the results obtained for observer's viewing angles of
50$^{\circ}$ and 90$^{\circ}$ using equation~(\ref{eq:pol}), it is likely
that the angle between the radio axis and the line of sight for 0850--206
is close to 90$^{\circ}$, given that for angles below 90$^{\circ}$ the
resulting unpolarized contribution is negative (particularly in the
1700\,--\,1850\,\AA \ range). This orientation is also consistent with the
radio image of this galaxy (see Fig.~\ref{vradalpha0850}).

On the other hand, in the case of 1303+091, the stellar population can
account for up to 50 per cent of the continuum emission if the radio axis
is on the plane of the sky, or its contribution could be null if the angle
between the line of sight and the radio axis is close to 50$^{\circ}$.

An alternative analysis consists of fitting the spectra of both galaxies
with four components: nebular continuum, scattered quasar light, young
stellar population and old stellar population. The quasar spectrum used
for the fits was a radio-loud quasar composite (from \pcite{telfer2002};
with a power-law index $\alpha_{\nu}$\,=\,--0.67 for rest-frame
wavelenghts $\lambda$\,$\gtrsim$\,1300\,\AA, and derived using over 200
spectra taken with the Hubble Space Telescope, with redshifts
z\,$>$\,0.33).  The quasar spectrum was then scattered using
\scite{zubko2000} dust scattering models, both optically thin and
optically thick (see Section~\ref{diss:scatterers}). Both old and young
stellar populations were modelled using the 2000 version of the GISSEL
codes of Bruzual \& Charlot (cf. \pcite{bruzual93},
2003\nocite{bruzual2003}), assuming a Salpeter IMF, solar metallicity and
instantaneous burst models of age 4 Gyr and 10 Myr (approximate age of the
radio source), respectively.  The nebular continuum contribution was fixed
by our observations and the other three components were allowed to
vary. Another variable was included in the fit to account for any
plausible intrinsic reddening in each galaxy.

The result of the fits is shown in Fig.~\ref{sedfit}. The best fit in both
cases was obtained when the quasar spectrum had been scattered by
optically thin dust, and for both galaxies a small intrinsic reddening was
required: E$_{\rm B-V}^{\rm intrin}$\,=\,0.087 for 0850--206, E$_{\rm
B-V}^{\rm intrin}$\,=\,0.020 for 1303+091. This intrinsic reddening could
plausibly be due to the same dust that is causing the scattering. The
contribution of old stellar population in the best-fitting model is
consistent at K-band wavelengths with typical K-band magnitudes for
z\,$\sim$\,1.4 radio galaxies (see K-z relation; \pcite{mccarthy93}). The
fits are good in general apart from the wavelength region $\sim$
2000\,--\,2200\,\AA \ where the fit underpredicts the observed continuum
emission (for further discussion see Section~\ref{diss:scatterers}).  The
results from the fitting show that in 0850--206 the scattered quasar light
dominates the UV continuum emission and there is no requirement at all for
a young stellar component; whilst in the case of 1303+091, both scattered
quasar light and young stellar population have important contributions to
the UV light: in the wavelength range 1700\,--\,1850\,\AA \ contributions
are comparable, while in the 2900\,--\,3050\,\AA \ range scattered quasar
light contributes more than 50 per cent. These results are in complete
agreement with our calculation above (see Table~\ref{poltabboth}).

Characteristic photospheric absorption features of OB stars, such as
SV\,1502 and NIV\,1720, were searched for in the spectra of both sources,
especially in 1303+091, where the contribution of a young stellar
population is suspected to be significant. No clear evidence for any of
these absorption lines was found in any of the spectra, although the S/N
of the spectra is insufficient for their absence to rule out a young
stellar population.

\subsubsection{The nature of the scatterers}
\label{diss:scatterers}

The nature of the scattering material in high-redshift radio galaxies is
an important question, still under debate.  Both electrons \cite{fabian89}
and dust grains (Tadhunter et al. 1988) have been proposed as possible
scattering agents; but dust is generally preferred, mainly because of the
greater scattering efficiency of dust grains over electrons.  Polarization
observations of radio galaxies show that for the implied scattered light
to be fully scattered by electrons, the required mass of ionized gas is
very large ($\sim$\,10$^{10}$\,--\,10$^{12}$\,M$_{\sun}$;
e.g. \pcite{di-serego94,cimatti96}). Thus, although some light might be
scattered by electrons (mainly in the central regions), the scattering
process is likely to be dominated by the dust grains (provided that some
dust is present).
 
\begin{figure*}
\centerline{ \psfig{figure=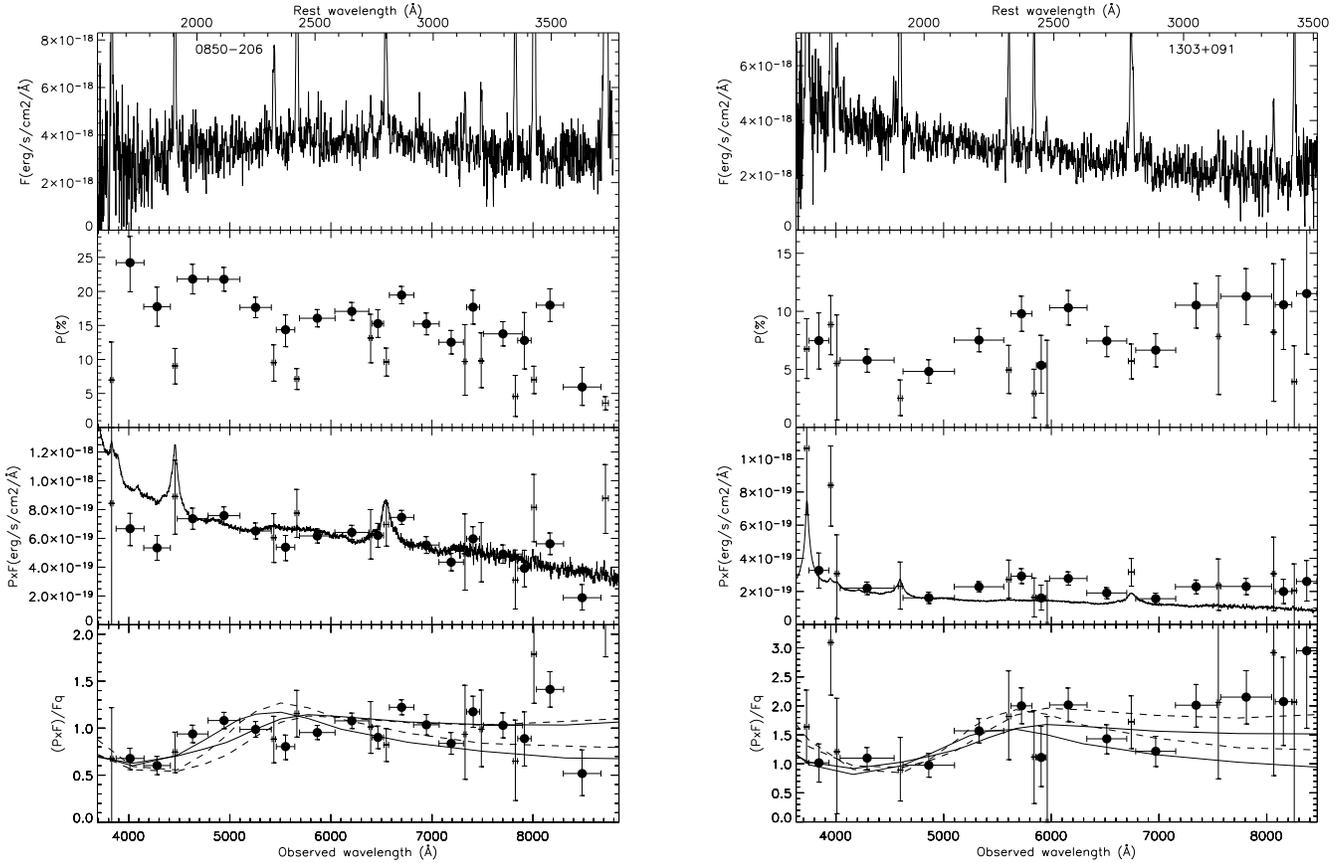,width=17.7cm}}
\caption[]{Polarization properties of 0850--206 (left) and 1303+091
(right). For each panel, from top to bottom: the unsmoothed total flux
spectrum in an expanded scale to show the continuum, the observed
fractional polarization, the polarized flux with a composite radio-loud
quasar spectrum [\,from \scite{telfer2002}, normalized to the polarized
flux at rest-frame 2000\,\AA] superimposed, and the ratio between the
polarized flux spectrum and the composite quasar spectrum with the
\scite{zubko2000} models overplotted [\,for optically-thin (thin lines)
and optically-thick (thick lines) dust cases, with scattering angles of
90$^{\circ}$ (solid lines) and 120$^{\circ}$ (dashed lines)]. Filled
circles and crosses indicate respectively continuum bins and emission line
bins with their underlying continuum.}
\label{polflux}
\end{figure*}

\scite{zubko2000} present detailed calculations of the scattering and
polarization properties of light scattered by both optically thin and
optically thick dust, using the \scite{mathis77} Galactic dust model. They
find that the spectrum of the scattered light is quite different in the
two cases. In the optically thin case the emergent scattered spectrum is
bluer than the incident one for $\lambda$\,$>$\,2200\,\AA, and it has a
broad dip centred at $\lambda$\,$\sim$\,1500\,\AA. In the optically thick
case the emergent spectrum is generally gray scattered for
$\lambda$\,$>$\,2200\,\AA, and shows a drop at
$\lambda$\,$<$\,2200\,\AA. Although the total scattered intensity depends
on optical depth, the polarization of the scattered light varies with
wavelength in a very similar way for both optically thin and optically thick
cases. The polarization tends to decrease slightly from 10000\,\AA \ to
$\sim$\,2000\,\AA, shows a dip centred at $\sim$\,1800\,\AA, and then
tends to increase sharply towards the far-UV.

In the case of quasar light scattered by electrons, the emergent spectrum
will be similar to the incident one, since the Thomson scattering
cross-section is independent of wavelength. Thus, for
$\lambda$\,$>$\,2200\,\AA, optically thick dust scattering can mimic
electron scattering.

In Fig.~\ref{polflux} continuum polarization properties of 0850--206
(left) and 1303+091 (right) are presented. In each panel, from top to
bottom the following are plotted: the total flux spectrum, the observed
fractional polarization, the polarized flux with a composite radio-loud
quasar spectrum (from \pcite{telfer2002}) superimposed, and the ratio
between the polarized flux spectrum and the composite quasar spectrum. The
dust-scattering predictions from \scite{zubko2000} are overplotted on this
bottom plot (see caption for details).

This composite quasar spectrum can be considered to be an average quasar
spectrum, and thus it can be compared with the polarized flux spectra of
0850--206 and 1303+091 (Fig.~\ref{polflux}). It is found that, for both
galaxies, the polarized spectrum is slightly redder than the incident one,
which is in agreement with spectropolarimetric studies of other
high-redshift radio galaxies (e.g. \pcite{cimatti96,kishimoto2001}).  The
ratio between polarized and incident spectra can be directly compared with
the scattering efficiency $\times$ intrinsic polarization
[\,cf. equation~(\ref{eq:pol})]; the dust scattering predictions of
\scite{zubko2000} were used for this comparison. It can be seen that the
broad dip shortward of $\sim$\,2200\,\AA \ (rest-frame) predicted by both
optically-thin and optically-thick dust models is found in both galaxies,
more pronounced in 1303+091.  It therefore appears that dust dominates the
scattering process in both 0850--206 and 1303+091.

The detection of features in the scattered light spectrum around 2200\,\AA
\ is particularly interesting when compared to the extinction laws of
different galaxies. The extinction curve of the Milky Way shows a strong
2200\,\AA \ dust feature; this feature is less pronounced for the Large
Magellanic Cloud (LMC), is even weaker for the Small Magellanic Cloud
(SMC), and is entirely absent in nearby starburst galaxies
(e.g. \pcite{calzetti94}). These differences are likely to be related to
the metallicity of the galaxy and the composition of the dust. The fact
that a 2200\,\AA \ dust feature is detected in the total flux spectra
(Fig.~\ref{sedfit}) and even more strongly in the polarized flux spectra
(Fig.~\ref{polflux}) of the two radio galaxies studied in this paper,
implies that the dust in these distant galaxies is different in nature or
composition to that of nearby starburst galaxies, and is much more
comparable to Galactic dust. On the other hand, the fits to the total
continuum spectra (Fig.~\ref{sedfit}) show that, although the observed
continuum and the model have the same shape around in the wavelength range
$\sim$\,2000\,--\,2200\,\AA, the model (which uses a Galactic dust
scattering prediction) slightly underpredicts the continuum emission in
this range. This suggests that the 2200\,\AA \ feature is not as strong as
in the models, and that the dust in these galaxies, although similar, may
not be exactly the same as Galactic dust.

\subsection{Ionization of the emission-line regions}

\begin{figure}
\centerline{ \psfig{figure=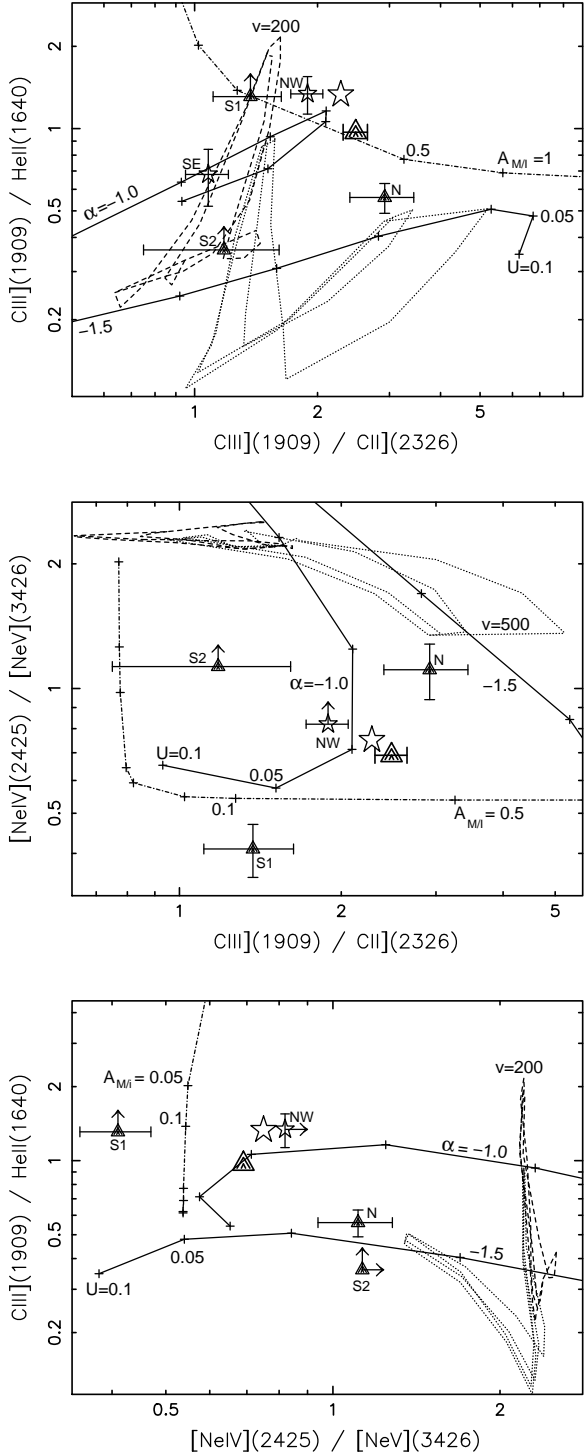,width=7.85cm}}
\caption[]{Diagnostic diagrams for different regions in 0850--206
(triangles) and 1303+091 (stars). Bigger symbols indicate the nucleus,
smaller ones denote EELR. Solid lines represent optically thick
single-slab power-law ($F_{\nu}$$\propto$$\nu^{\alpha}$) photoionization
models (MAPPINGS) with $\alpha$=--1.0,--1.5 and a sequence in ionization
parameter (2.5$\times$$10^{-3}$$<$U$<$$10^{-1}$). The dot-dash-dot line
corresponds to photoionization models including matter-bounded clouds
(10$^{-2}$$\leq$A$_{\rm M/I}$$\leq$10; \pcite{binette96}).  Shock (dashed
grid) and shock+precursor (dotted grid) models are included
\cite{dopita96}; each sequence corresponds to a fixed magnetic parameter
(B/$\sqrt{n}$=0,1,2,4 $\mu$Gcm$^{-3/2}$) and changing shock velocity
(200$\leq$$v$$\leq$500 \kms).}
\label{regdiag}
\end{figure}

In the above sections where the results are described, the spatial
variation of the line ratio CIII]1909\,/\,CII]2326 along the position of
the slit for 0850--206 and 1303+091 were presented. For 1303+091 it was
found that the ionization state peaks in the nuclear region, with a
possible second peak at $\sim$\,1.2 arcsec to the north-west of the
nucleus. For 0850--206 the highest ionization state was found $\sim$\,0.5
arcsec offset of the nucleus to the north, then it decreases at both sides
of this location.

In order to study in more detail the ionization state of the emission-line
gas in the two galaxies, the line ratios (derived from the total intensity
spectra) for the different regions have been compared with theoretical
predictions of several ionization models. The resulting diagnostic
diagrams are presented in Fig.~\ref{regdiag}, which involve the following
line ratios: CIII]1909\,/\,HeII\,1640, CIII]1909\,/\,CII]2326, and
[NeIV]2425\,/\,[NeV]3426. The points correspond to the different
emission-line regions defined along the slit (for details of the apertures
see captions of Figs.~\ref{0850specreg} and \ref{1303specreg}).

The data are compared with the following ionization models: (i)
optically-think power-law photoionization, with spectral indexes $\alpha=$
--1.0 and --1.5 (solid lines); (ii) photoionization including
matter-bounded clouds (dot-dash-dot line); (iii) shock-ionization, with
shock velocities in the range 200\,$\leq$\,$v$\,$\leq$\,500 km\,s$^{-1}$
(dashed grid); and (iv) shock-ionization including a photoionized
precursor (dotted grid). For details about how the different models were
generated, see Sol\'{o}rzano-I\~{n}arrea et al. (2001) and references
therein.

By comparing the three diagnostic diagrams, it can be seen that the points
of the nuclear regions (bigger symbols) of the two galaxies fall close
together in the three diagrams, indicating that they have a similar state
of ionization. Both nuclei are well reproduced by the photoionization
predictions with $\alpha$\,$\sim$\,--1.0 and U\,$\sim$\,0.025. Note that
they could also be explained by the mixed-medium photoionization models,
since these models can be slightly tuned to better fit the data. The
shock-ionization predictions, however, clearly fail to reproduce the
ionization state of the nuclear regions in the two galaxies.

The EELR of 1303+091 (open stars) are well explained by the
photoionization models with $\alpha$\,$\sim$\,--1.0, although the data can
also be consistent with low velocity shock-ionization predictions. Note,
however, that in the middle and lower diagrams there is no data for the SE
region, since none of the neon lines was measurable in this region and
therefore the ratio [NeIV]\,/\,[NeV] could not be calculated. Only a
lower limit could be derived for the same line ratio for the NW
region. Thus, the ionization of the EELR in 1303+091 cannot be well
constrained.

The EELR of 0850--206 (triangles) fall well apart from each other in the
diagrams. Whilst the ionization of the northern region N can be explained
by $\alpha$\,$\sim$\,--1.5 photoionization predictions and by
shock+precursor models with shock velocities of $\sim$\,500\,km\,s$^{-1}$;
the southern region S1 is best explained by mixed-medium photoionization
models with 0.05\,$<$\,A$_{\rm M/I}$\,$<$\,0.1. On the other hand, the
ionization of the outermost southern region S2 is poorly constrained and
thus can be reproduced by either the photoionization models or
shock-ionization predictions with or without a precursor gas.

In summary, it is found that the ionization state of the nuclear regions
in the two galaxies can only be reproduced by photoionization
predictions. The photoionization models give also reasonable fits to the
data of the extended regions. However, their ionization is less
constrained, and with the exception of the S1 region in 0850--206, whose
ionization is only consistent with the mixed-medium photoionization
models, the ionization of the other EELR could also be explained by
shock-ionization predictions. In particular, the SE region of 1303+091,
which coincides with the location of a radio knot, was found to have the
lowest ionization state and the highest emission linewidths; these
features are expected when the gas is perturbed by strong interactions
with the radio source structures (e.g. \pcite{clark98}).

\subsection{Variations with radio source size}

Clearly, continuum studies of only two sources are not enough to draw
significant conclusions about the evolution of the continuum properties of
radio galaxies with radio source age. Nevertheless, the results presented
in this paper are in line with what might have been expected. 

If a jet-induced starburst occurs when the radio hotspots pass through the
host galaxy, by a few $\times$\,10$^{7}$ years later (typical age of a
radio source) the starburst luminosity will have decreased to $\sim$\,10
per cent of its peak value \cite{best96}.  Therefore, the presence of a
young stellar population, possibly jet-induced, may dilute the scattered
AGN light in smaller (younger) radio sources and, consequently, the
observed continuum polarization would be lower. On the other hand, for
larger (older) radio sources, in which the contribution of a starburst to
the total optical flux will be dramatically decreased, the scattered AGN
light dominates the UV continuum emission.  Our findings are consistent
with this scenario. 0850--206, the larger radio source of the two (D$_{\rm
rad}$\,$\sim$\,118 kpc), presents a high continuum polarization, averaging
$\sim$\,17 per cent; scattered quasar light dominates the UV continuum
emission and the starlight contribution is null or negligible. By
contrast, 1303+091 (D$_{\rm rad}$\,$\sim$\,73 kpc) presents a lower
continuum polarization, averaging $\sim$\,8 per cent, and in this galaxy
the young stellar population could account for up to $\sim$\,50 per cent
of the UV continuum emission.

\begin{figure}
\centerline{ \psfig{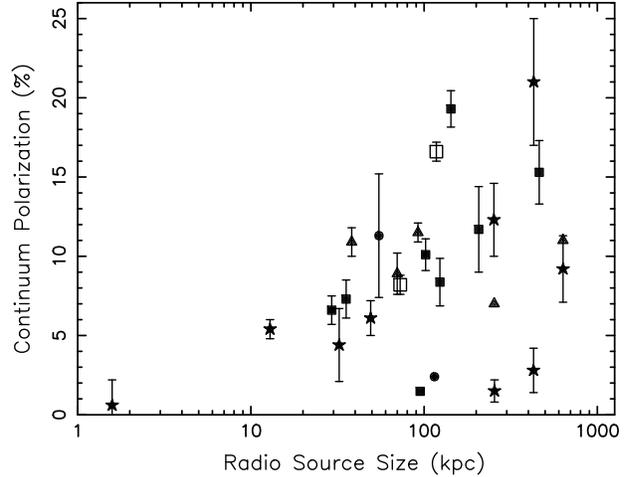}}
\caption[]{Observed fractional continuum polarization of high-redshift
radio galaxies versus their radio source size. Different symbols indicate
different redshift ranges: 0.5\,$<$\,z\,$<$\,1.0 (filled stars),
1.0\,$<$\,z\,$<$\,2.0 (filled triangles), 2.0\,$<$\,z\,$<$\,3.0 (filled
squares) and 3.0\,$<$\,z\,$<$\,4.0 (filled circles). Also plotted are the
two radio galaxies from our sample presented in this paper (big open
squares). The data corresponding to the other galaxies have been taken
from di Serego Alighieri et al. (1989, 1993, 1996); \scite{tadhunter92};
Cimatti et al. (1993, 1996, 1997); \scite{dey96,dey97} and
\scite{vernet2001}. Information about the sizes of radio sources has been
obtained from \scite{dunlop89,mccarthy91,rottgering94};
\scite{carilli94a,carilli97,cohen97} and \scite{kapahi98}.
\nocite{di-serego89,di-serego93,di-serego96,cimatti93,cimatti96,cimatti97}}
\label{polsize}
\end{figure}

These results can be combined with those of previous polarization studies
of high-redshift radio galaxies, in order to explore if there is any
correlation between continuum polarization and size of the radio
source. Fig.~\ref{polsize} shows the fractional continuum polarization
versus radio source size for a sample of high-redshift radio galaxies
taken from the literature (see caption of the figure for details).  It
must be emphasized that this set of radio galaxies consists of several
different subsamples, many of which are biased (e.g. towards high
polarization objects). In addition, these samples are at different
redshifts, making the wavelength range studied different and bringing
evolution effects into play, thus making the objects not directly
comparable.  However, even despite this, it is intriguing that there is a
hint of a weak correlation between polarization and radio source size
(Spearmann-Rank test gives a 97 per cent significance level): smaller
radio sources do not show high polarizations, and the larger the radio
source the higher its polarization can be.

The relation between radio source size and polarization will be better
tested when our full sample of radio galaxies at the same redshift
(z$\sim$1.4), with a large range of radio sizes, has been observed. Also,
it will be interesting to sum~the spectra of the smaller radio sources in
the full sample, with similar characteristics to 1303+091, to search at
higher S/N for photospheric absorption features of OB stars, which would
directly prove the presence of a young stellar population.

\section{Summary and conclusions}

This paper presents the results obtained from VLT spectropolarimetric
observations of two powerful radio galaxies at z\,$\sim$\,1.4. Analysis of
their scattered flux spectra and their polarization properties indicates
that the scattering process in these galaxies is dominated by dust, whose
2200\,\AA \ feature suggests that it is similar to Galactic dust in
nature. The larger radio source (0850--206) presents a continuum
fractional polarization reaching as high as $\sim$\,24 per cent at
$\sim$\,2000\,\AA \ (rest-frame). For this galaxy, scattered AGN light
must dominate the UV continuum, the nebular emission contributes up to
$\sim$\,22 per cent and there is no requirement for any young starlight
contribution. The smaller radio source (1303+091) presents a lower
continuum polarization averaging $\sim$\,8 per cent across the observed
wavelength range, and its nebular continuum contribution to the UV
continuum is $\sim$\,11 per cent. For this galaxy, the starlight could
contribute between 0 and 50 per cent, depending on the radio axis
orientation, with multi-component fit to the spectral energy distribution
favouring the upper end of this range.  In both galaxies, the position
angle of the electric vector is almost independent of wavelength and
within 15$^{\circ}$ of perpendicular to the radio axis.

The emission-line properties of the galaxies are also analysed.
Illumination by the central AGN is likely to be the dominant ionization
mechanism of the nuclear and EELR in both galaxies, although the region
coincident with the radio knot in 1303+091 shows features reminiscent of
jet-cloud interactions.

A compilation of polarimetric observations from the literature shows hints
of a correlation between continuum polarization and radio source size,
albeit for incomplete and biased samples. This is broadly in line with
what might have been expected; that in smaller radio sources the presence
of a young stellar population (possibly jet-induced) may dilute the
scattered AGN light, while in larger (older) radio sources the presence of
young stars is negligible and the scattered AGN light dominates the UV
continuum. It is clearly necessary to observe and analyse the rest of our
sample to confirm this result with a complete and unbiased sample, in
order to allow significant conclusions to be drawn about the evolution of
the continuum properties of radio galaxies with radio source size.

\section*{Acknowledgments}
We are very grateful to Joel Vernet for kindly providing his IDL
spectropolarimetric reduction code and to Carlos De Breuck for providing
his modified version of the code and for a constructive discussion about
reduction of VLT spectropolarimetric data. We are also indebted to Viktor
Zubko for kindly supplying us with the results of their dust scattering
models in digitised form. CSI thanks Makoto Kishimoto for useful
discussions. We also thank the anonymous referee for a careful reading of
the manuscript and useful comments that helped to improve the paper. This
work is based on observations made at the European Southern Observatory,
Paranal, Chile. This research has made use of the NASA/IPAC Extragalactic
Database (NED) which is operated by the Jet Propulsion Laboratory,
California Institute of Technology, under contract with the National
Aeronautics and Space Administration. The authors acknowledge the data
analysis facilities provided by the Starlink Project which is run by CCLRC
on behalf of PPARC. This research has also made use of the USNOFS Image
and Catalogue Archive operated by the United States Naval Observatory,
Flagstaff Station (http://www.nofs.navy.mil/data/fchpix/).

\bibliographystyle{mnras}
\bibliography{reference}

\end{document}